\documentclass{article}

\usepackage[accepted]{mlsys2023}
\usepackage{microtype}

\usepackage[utf8]{inputenc}
\usepackage{amssymb}
\usepackage{amsmath}
\usepackage{multirow}
\usepackage{chngcntr}
\usepackage{xcolor}
\usepackage{hyperref}
\usepackage{graphicx}
\usepackage{booktabs}
\usepackage{pifont}
\usepackage{xspace}
\usepackage{subcaption}
\newcommand{\cmark}{\ding{51}}
\newcommand{\xmark}{\ding{55}}
\usepackage{fontawesome}
\newcommand{\tu}{\faThumbsUp}
\newcommand{\td}{\faThumbsODown}

\counterwithin{table}{section}

\newcommand{\ddp}{\text{DP}\xspace}
\newcommand{\tp}{\text{TP}\xspace}
\newcommand{\pp}{\text{PP}\xspace}

\newcommand{\nsdp}{\ensuremath{\text{DP}_0}\xspace}
\newcommand{\psdp}{\ensuremath{\text{DP}_\text{PS}}\xspace}
\newcommand{\fsdp}{\ensuremath{\text{DP}_\text{FS}}\xspace}

\newcommand{\nsdpi}{0\xspace}
\newcommand{\psdpi}{\text{PS}\xspace}
\newcommand{\fsdpi}{\text{FS}\xspace}

\newcommand{\nlbf}{\ensuremath{\text{PP}_\text{gpipe}}\xspace}
\newcommand{\nldf}{\ensuremath{\text{PP}_\text{1f1b}}\xspace}
\newcommand{\ldf}{\ensuremath{\text{PP}_\text{DF}}\xspace}
\newcommand{\lbf}{\ensuremath{\text{PP}_\text{BF}}\xspace}

\newcommand{\ndp}{\ensuremath{N_\text{DP}}\xspace}
\newcommand{\ntp}{\ensuremath{N_\text{TP}}\xspace}
\newcommand{\npp}{\ensuremath{N_\text{PP}}\xspace}

\newcommand{\ngpu}{\ensuremath{N_\text{GPU}}\xspace}
\newcommand{\nnode}{\ensuremath{N_\text{Node}}\xspace}
\newcommand{\snode}{\ensuremath{S_\text{Node}}\xspace}

\newcommand{\bs}{\ensuremath{B}\xspace}
\newcommand{\mbs}{\ensuremath{S_\text{mb}}\xspace}
\newcommand{\mbc}{\ensuremath{N_\text{mb}}\xspace}
\newcommand{\seq}{\ensuremath{S_\text{seq}}\xspace}
\newcommand{\voc}{\ensuremath{S_\text{voc}}\xspace}

\newcommand{\nl}{\ensuremath{N_\text{layers}}\xspace}
\newcommand{\hid}{\ensuremath{S_\text{hidden}}\xspace}
\newcommand{\shead}{\ensuremath{S_\text{head}}\xspace}
\newcommand{\nhead}{\ensuremath{N_\text{heads}}\xspace}
\newcommand{\mlp}{\ensuremath{S_\text{mlp}}\xspace}
\newcommand{\nparams}{\ensuremath{N_\text{params}}\xspace}

\newcommand{\nstage}{\ensuremath{N_\text{stage}}\xspace}
\newcommand{\nloop}{\ensuremath{N_\text{loop}}\xspace}
\newcommand{\chimera}{\ensuremath{N_\text{Ch}}\xspace}

\newcommand{\tnet}{\ensuremath{T_\text{net}}\xspace}
\newcommand{\tcomp}{\ensuremath{T_\text{comp}}\xspace}
\newcommand{\tover}{\ensuremath{T_\text{overlap}}\xspace}

\newcommand{\bcrit}{\ensuremath{B_\text{crit}}\xspace}
\newcommand{\bmin}{\ensuremath{\beta_\text{min}}\xspace}
\newcommand{\btmin}{\ensuremath{\beta_\text{net}}\xspace}
\newcommand{\bgpu}{\ensuremath{\beta}\xspace}

\def\app#1#2{%
  \mathrel{%
    \setbox0=\hbox{$#1\sim$}%
    \setbox2=\hbox{%
      \rlap{\hbox{$#1\propto$}}%
      \lower1.1\ht0\box0%
    }%
    \raise0.25\ht2\box2%
  }%
}

\mlsystitlerunning{Breadth-First Pipeline Parallelism}

\begin{document}

\twocolumn[
\mlsystitle{Breadth-First Pipeline Parallelism}



\mlsyssetsymbol{equal}{*}

\begin{mlsysauthorlist}
\mlsysauthor{Joel Lamy-Poirier}{snow}
\end{mlsysauthorlist}

\mlsysaffiliation{snow}{ServiceNow Research, Montreal, Qu'ebec, Canada}

\mlsyscorrespondingauthor{Joel Lamy-Poirier}{joel.lamy-poirier@servicenow.com}

\mlsyskeywords{Machine Learning, MLSys}

\vskip 0.3in

\begin{abstract}
    We introduce Breadth-First Pipeline Parallelism, a novel training schedule which optimizes the combination of pipeline and data parallelism. Breadth-First Pipeline Parallelism lowers training time, cost and memory usage by combining a high GPU utilization with a small batch size per GPU, and by making use of fully sharded data parallelism. Experimentally, we observed an increase of up to 43\% in training throughput for a 52 billion-parameter model using a small batch size per GPU compared to Megatron-LM, which would reduce the training time and cost by the same amount on a large GPU cluster.
\end{abstract}
]

\printAffiliationsAndNotice{}

\section{Introduction}
\label{sec:intro}

Large language models \cite{transformer,openAiGpt3} are quickly becoming an essential tool for natural language processing. However, a challenging aspect of developing such models is their long and expensive training process. A single training may require tens, or even hundreds of thousands of GPU-days worth of computation \cite{openAiGpt3,nvidiaMegatron2,googleChinchilla}. This results in price tags that can reach several million dollars and a large environmental footprint. Significant efforts have been made towards reducing the training duration and cost, for example by improving the model \cite{googleSwitch}, the training scheme \cite{googleChinchilla,openAiScaling} or the hardware utilization \cite{nvidiaMegatron2,googlePalm,microsoftZero, nvidiaMegatron1,nvidiaMegatron3, flashAttention}. 
However, the training time and cost can only be jointly optimized up to a certain point, as there is an inherent trade-off between them. This trade-off is largely invisible for small models but becomes a limiting factor for large models with tens or hundreds of billions of parameters, that need to be trained on large GPU clusters.

On the one hand, reducing training time requires an increased number of GPUs (\ngpu), which in turn needs a larger batch size (\bs). These extra GPUs will typically be added through data parallelism, so they need to process \emph{different} samples. In general, distributed training requires a \emph{minimum batch size per GPU} (\bmin), which is typically equal or slightly smaller than one.
In practice, most models are trained with a batch size per GPU much higher than this bare minimum, to allow for a higher GPU utilization.

On the other hand, increasing the batch size hurts the effectiveness of stochastic gradient descent (SGD). The efficiency is maximal when the batch size is well-below an empirical value known as the \emph{critical batch size} $\bcrit$ \cite{criticalBatch}, $\bs\ll\bcrit$.
However, this \emph{small batch size} regime is unattainable on larger clusters since $\bs\geq\bmin\ngpu$.
A large body of work \cite{criticalBatch, openAiScaling,shallue2018,Goyal2017AccurateLM,Smith2018DontDT} has demonstrated that larger batches are able to train machine learning models given a careful adjustment of the training hyperparameters, but they slow down training by requiring extra training samples to reach the same validation loss. That is, they add an overhead which increases the training cost (and time). 


\begin{figure}[t]
    \centering
    \subcaptionbox{
        Training time
        \label{fig:bert_52b_4096gpus_durations}
    }{
        \includegraphics{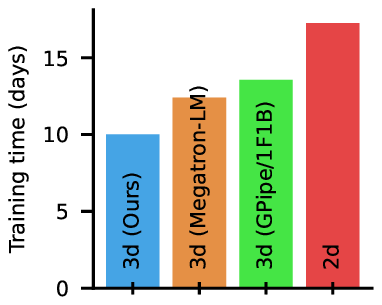}
    }\hspace{-8pt}\subcaptionbox{
        Memory
        \label{fig:bert_52b_4096gpus_memories}
    }{
        \includegraphics{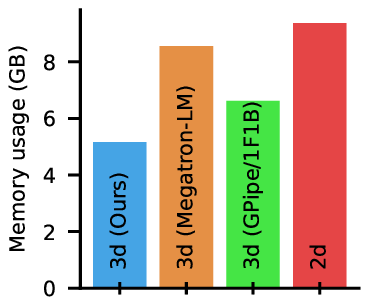}
    }
\caption{Predicted training time (a) and memory usage (b) for a 52 billion parameter model on a cluster of 4096 Nvidia V100 GPUs, using our method (Breadth-First Pipeline Parallelism), compared to 3d and 2d baselines. 
}
\label{fig:speedup_52b_4096_gpus}
\end{figure}

Thus, the trade-off can be summarized as follows: reducing the training time requires a larger batch, but a large batch increases the cost and has diminishing returns beyond a certain point. We stress that this concerns the entire training process rather than the batch time or GPU utilization which, while important, do not tell the full story.
Although this trade-off is difficult (if not impossible) to avoid, we can mitigate it by reducing the batch size per GPU as much as possible, ideally to $\beta_\text{min}$. However, there is a major obstacle to doing so: existing parallelization methods are inefficient in this \emph{small batch size per GPU} regime. Indeed, the state-of-the-art methods such as 2d \cite{googlePalm, facebookOpt} and 3d \cite{nvidiaMegatron2, nvidiaMegatron3} parallelism are able to achieve a high GPU utilization (i.e., to use a high fraction of the available flop/s) for a wide range of model sizes, but require a batch size per GPU significantly higher than \bmin to do so.

Therefore, to train large language models more efficiently, we should look for a training method that not only achieves a high GPU utilization, but that does so with a \emph{low batch-size per GPU}. We propose a novel method, Breadth-First Pipeline Parallelism, that achieves precisely that by using a \emph{looping} placement of the network layers, together with a \emph{breadth-first} micro-batch schedule. Looping pipelines provide a way to reduce the pipeline-parallel overhead from the \emph{pipeline bubble}, as opposed to the more common mitigation method of increasing the batch size. They were first introduced in \cite{nvidiaMegatron2}, where they allowed for an increased computational efficiency. However, we show that the \emph{depth-first} schedule used in that paper is sub-optimal for two main reasons. First, it increases the network overhead, which negates much of the benefit from looping. Second, looping pipelines are more efficient for \emph{smaller} pipelines, which for larger models are prevented by memory constraints. The breadth-first schedule avoids both of these limitations: it allows lowering the memory usage to a minimum with \emph{fully sharded data parallelism} \cite{microsoftZero} and prevents a network bottleneck by maximizing the \emph{overlap} between network communication and computation. Experimentally, we observed an increase in throughput of up to 43\% near $\beta_\text{min}$ for a 52 billion parameter model, which translates into a similar (though slightly lower) reduction in training cost and time reduction on large clusters (Figure \ref{fig:speedup_52b_4096_gpus}).


This paper is organized as follow. In section \ref{sec:extended_intro}, we clarify our main claim and its assumptions. In section \ref{sec:background}, we introduce the required background on distributed training. In section \ref{sec:breadth_first_pipeline}, we introduce our main method, Breadth-First Pipeline Parallelism, and summarize its theoretical justification. In section \ref{sec:evaluation}, we demonstrate our claims experimentally.

\section{Extended introduction}
\label{sec:extended_intro}

Our contribution is summarized as follows: Breadth-First Pipeline Parallelism reduces the time and/or cost of training large language models on large GPU clusters, when compared to state-of-the-art methods. Before continuing, we clarify the meaning of this claim and its assumptions.

\paragraph{Time and cost} We assume the time and cost to be important for obvious reasons. There may be some flexibility on the training time, but we assume that a \emph{reasonable} training time does not exceed a few months. The price tag depends on many factors, but we approximate it by the total hardware usage in GPU-days. We assume that the type of hardware used is outside of our control, so the training cost and time are determined by the total compute requirement, number of GPUs used and the \emph{GPU utilization}, defined as the fraction of the available computing power that is effectively used:
\begin{equation}
    \text{Cost}\propto\frac{\text{Total compute}}{\text{Utilization}},\quad
    \text{Time}\propto\frac{\text{Cost}}{\text{Num GPUs}}.
    \label{eq:time_cost}
\end{equation}
In this paper we are particularly interested in the impact of the number of GPUs, which both the total compute and utilization may depend on.

\paragraph{Large GPU cluster} Our method is aimed at clusters with hundreds or thousands of GPUs, for which the batch size is a limiting factor due to its effect on SGD (Section \ref{subsec:critical_batch}). We assume a cluster of modern NVIDIA GPUs such as V100s or A100. Such a cluster normally consists of several \emph{nodes} (servers) with several GPUs (typically 8) connected with a very fast NVLink connection. The nodes are connected via a slower InfiniBand network. Other types of clusters, such as TPU pods, should also benefit from our method but may affect certain aspect of our analysis, especially when the network structure is different.

\paragraph{Training} This refers to pre-training. Fine-tuning may also benefit from our method, but it typically runs on small clusters for which the batch size per GPU is not as important.

\paragraph{Large language model} Large language models, i.e. models with a transformer architecture \cite{transformer} and more than a few billion parameters, are the main use case for our method, largely because of their size and computational requirement. Smaller models should also benefit from our method but may invalidate certain aspects of our analysis. Other architectures are also possible under certain assumptions, most importantly they should admit a breakdown into similarly sized layers for efficient pipeline parallelism.

\paragraph{State-of-the-art} By state-of-the-art, we principally refer to methods that were successfully used to train large language models. These methods all consists of a combination of (up to) three basic methods: \emph{data parallelism} (\ddp), \emph{pipeline parallelism} (\pp) and \emph{tensor parallelism} (\ddp). We describe each of this method and their variations in the next section. The state-of-the-art are \emph{2d and 3d parallelism}, as described in \ref{subsec:combined_methods}. We exclude the methods introduced in \cite{nvidiaMegatron3} (\emph{sequence parallelism} and \emph{selective activation recomputation}), as the paper was published after our codebase was completed. However, these methods are largely orthogonal to ours so should work well in combination. In fact, the lower memory usage of Breadth-First Pipeline Parallelism should make it easier to avoid recomputing activations. 

\section{Distributed training}
\label{sec:background}

In this section, we review the three basic methods (\ddp, \pp and \tp), both in isolation and in combination with others. We place emphasis on the memory usage from the training state (weights, optimizer momenta), which for large models largely exceeds the memory available on a single GPU, and on the \emph{batch size per GPU} ($\bgpu=B/\ngpu$). We also take a special look at the interaction between data and pipeline parallelism, which is the focus of the present paper.
Finally, we review how the batch size impacts the training process, which effectively sets a limit on distributed scaling. 
We remain qualitative and refer to the appendix for more detailed results and examples.

All distributed methods involve network operations, and for efficient training these operations should have a minimal overhead. To achieve this, the operation may be \emph{overlapped} (run in parallel) with computation.\footnote{For simplicity, we assume that the overlap is perfect, i.e. that it does not add any overhead on either operation. In practice, there may be a small overhead; for example, on a Nvidia A100 GPU, the InfiniBand network transfers uses 2 of the 108 execution units (\emph{SM}), slowing the computation by approximately 2\%.} An overlapped network operation has a negligible overhead provided that its duration (\tnet) is less than that of the overlapped computation (\tover). If overlap is impossible, the network operation should instead be short compared to the total computation time (\tcomp). In short, efficiency requires either
\begin{equation}
    \label{eq:efficiency}
    \tnet\leq\tover \quad\text{or}\quad \tnet\ll\tcomp.
\end{equation}

\subsection{Data parallelism}
\label{subsec:data}

Data parallelism (\ddp) divides the batch between the \ndp devices. Each device calculates the loss and gradients for its input, then shares its results through \emph{gradient reduction} and updates the weights. 

The input consists of \ndp parallel and \mbc sequential micro-batches of size \mbs, for a batch size $\bs=\ndp\mbc\mbs$. With pure \ddp, the batch size per GPU $\bgpu=\mbc\mbs$ is at least one, i.e., there is a \emph{minimum batch size per GPU} $\bmin=1$. However, training at \bmin may be inefficient. First, a higher micro-batch size leads to more efficient computational kernels due to increased \emph{thread-level parallelism} and reduced \emph{memory IO overhead}, though this is mainly relevant for smaller models, and larger ones generally allow for efficient kernels at any micro-batch size. Second, the gradient reduction is generally a bottleneck at \bmin. With data parallelism alone, the gradient reduction time is fixed, so additional computation is needed to satisfy Eq. (\ref{eq:efficiency}), i.e., a higher batch size. The computation time is approximately proportional to the amount of computation, hence the batch size. However, only one of the sequential micro-batches can be overlapped, so the overlapped time is proportional to the micro-batch size. Summing up, we can rewrite Eq. (\ref{eq:efficiency}) as\footnote{This can be derived by substituting $\tnet\propto1$, $\tcomp\propto\mbc\mbs=\bgpu$ and $\tover\propto\mbs=\tfrac\bgpu\mbc$, and by absorbing the proportionality factor into a constant \btmin.}
\begin{equation}
    \bgpu\geq\mbc\btmin \quad\text{or}\quad \bgpu\gg\btmin
    \label{eq:ddp_efficiency}
\end{equation}
for some constant \btmin.  Intuitively, \btmin represents the lowest value of \bgpu for which Eq. (\ref{eq:efficiency}) can be satisfied. Its exact value depends on the hardware, model and implementation, but is almost always larger than one. As an example, OPT-175B \cite{facebookOpt} was trained with a micro-batch size of 8, which suggests $\btmin\lessapprox8$ for that setup. Note that in the overlapped case, \btmin is effectively a strict threshold because there is a sharp decline in training efficiency below this value (Figure \ref{fig:vary_batch_size_theory_overlapped}).

In the original form of data parallelism (\nsdp), the computed gradients are all-reduced (summed) between the devices, after which the weights are updated redundantly on each of them. However, \nsdp is inefficient from a memory perspective as it requires a duplication of the whole training state on every device. This duplication can be avoided with \emph{partially sharded data parallelism} (\psdp) \cite{microsoftZero}, where each device instead optimizes a fraction (\emph{shard}) of the weights. The weights are reduce-scattered on the appropriate devices, then updated and \emph{reconstructed} (all-gathered) back on all devices. Due to the efficiency of the network operations, the communication volume remains the same as with \nsdp.
Given enough data parallelism, \psdp divides the memory usage from the training state by up to 8 times (see Appendix \ref{subsubsec:memory_state}). However, this reduction may still not be sufficient for very large models.

The memory usage can be reduced further with \emph{fully sharded data parallelism} (\fsdp),\footnote{In the language of \cite{microsoftZero}, \psdp corresponds to stage two, while \fsdp below corresponds to stage three.} where the layers are not kept on device and are instead reconstructed prior to every use. Each layer is reconstructed in both the forward and backward passes, increasing the network usage by at least 50\%. The network operations are also repeated for every micro-batch,\footnote{This can be avoided with the breadth-first schedule introduced in Appendix \ref{sec:breadth_first_2d}.} so the usage is also multiplied by \mbc. In short, \fsdp shrinks the memory usage from the layer weights and gradients to a minimum (typically that of two layers), but increases the network usage, especially with gradient accumulation.

Data parallelism alone can be used to train large models, with \fsdp. However, it requires a high batch size per GPU, which makes it less efficient on large clusters (see Section \ref{subsec:critical_batch}). Scaling can be improved by combining with model parallelism (pipeline or tensor), to which we now turn.


\subsection{Pipeline parallelism}
\label{subsec:pipeline}

Pipeline parallelism (\pp) is a form of \emph{model parallelism}, dividing the model along its \emph{depth} \cite{gpipe}. Each of the \npp pipeline-parallel devices hosts a single contiguous set of layers, or a \emph{stage} (Figure \ref{fig:line_placement}). In particular, it only stores a fraction of the training state memory. The stages should be identical or near identical in size, so that they take about the same time (and memory) to process a micro-batch.

Parallel computation is achieved with multiple (\mbc) sequential micro-batches, with $\mbc\geq\npp$ ($\bmin=1$) so that all devices may perform computation at the same time. However, the data takes time to traverse the pipeline, which causes the devices to be idle (input-starved) much of the time. This phenomenon, known as the \emph{pipeline bubble}, adds an overhead equivalent to $\npp-1$ micro-batches, or
\begin{equation}
    \text{Bubble}=\frac{\npp-1}{\mbc}.
    \label{eq:bubble}
\end{equation}
Therefore, $\mbc\gg\npp$ is required for computational efficiency. Although this is a worse requirement than for \ddp, the method \emph{does} allow for training with a lower batch size, at a reduced efficiency (Figure \ref{fig:vary_batch_size_theory_overlapped}, non-looped). The bubble is maximal at \bmin, with an overhead of up to 100\%.

When compared to the other methods \pp requires the lowest amount of network communication, which is mostly negligible for the large models and fast networks considered in this paper. This communication can also be overlapped with computation, which requires $\mbc\geq\npp+1$ since a micro-batch cannot take part in computation while being transferred.

\begin{figure*}[tp]
    \centering
    \begin{subfigure}{\textwidth}
    \centering
        \includegraphics{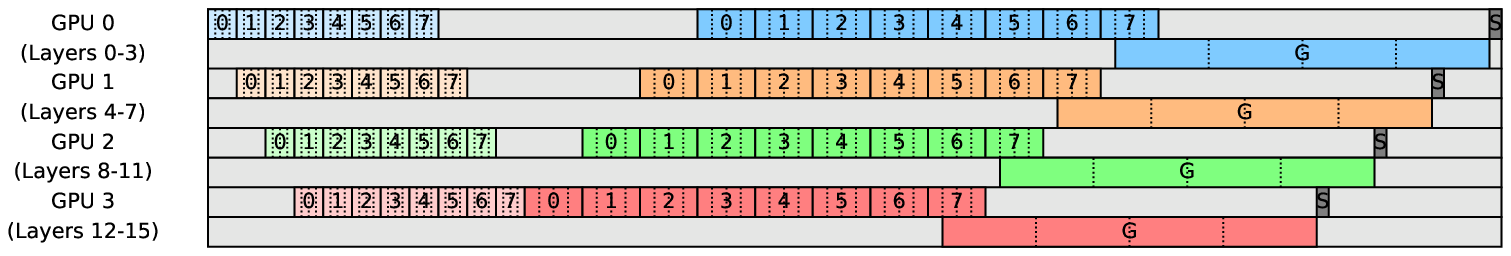}
        \includegraphics{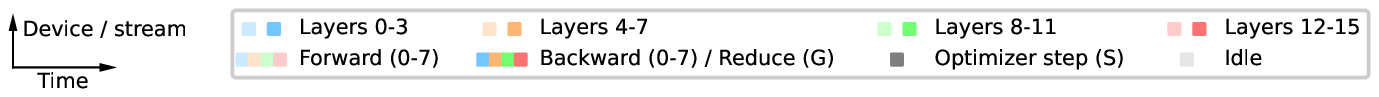}
        \caption{Non-looped pipeline, GPipe schedule (\nlbf): large bubble, poor overlap}
        \label{fig:schedule_gpipe}
    \end{subfigure}\vspace{8pt}
    \begin{subfigure}{\textwidth}
    \centering
        \includegraphics{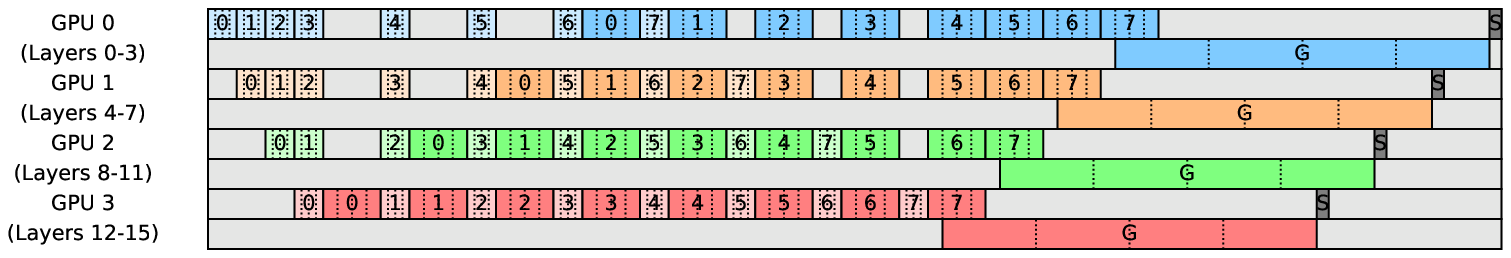}
        \includegraphics{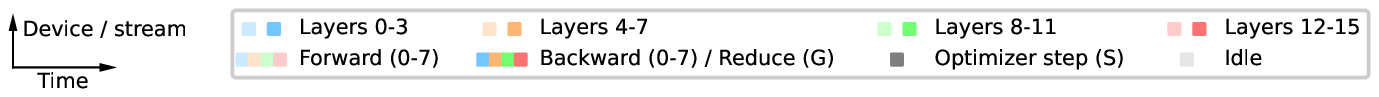}
        \caption{Non-looped pipeline, 1F1B schedule (\nldf): large bubble, poor overlap}
        \label{fig:schedule_1f1b}
    \end{subfigure}\vspace{8pt}
    \begin{subfigure}{\textwidth}
    \centering
        \includegraphics{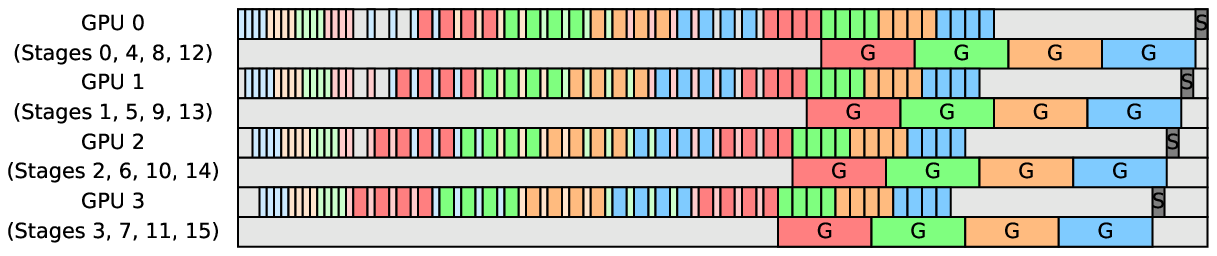}\\
        \includegraphics{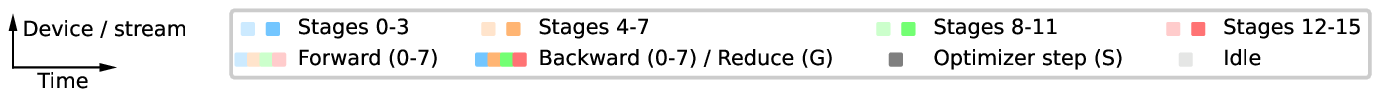}
        \caption{Looped pipeline, depth-first schedule (\ldf): small bubble, moderate overlap}
        \label{fig:schedule_depth_first}
    \end{subfigure}\vspace{8pt}
    \begin{subfigure}{\textwidth}
    \centering
        \includegraphics{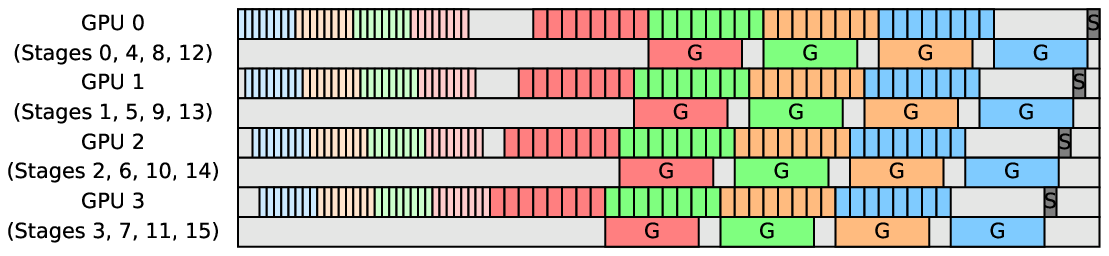}\\
        \includegraphics{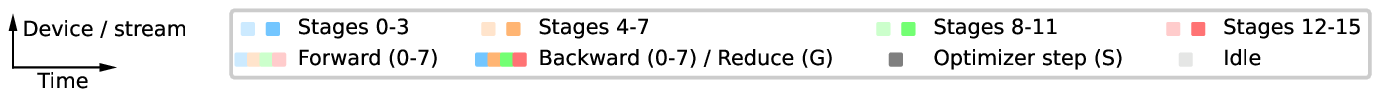}
        \caption{Looped pipeline, breadth-first schedule (\lbf): small bubble, best overlap}
        \label{fig:schedule_breadth_first}
    \end{subfigure}
    \caption{
    Comparison of the four pipeline schedules considered in this paper (times to scale), for a 16-layer model on 4 pipeline devices, with 8 sequential micro-batches (numbered 0-7), in the presence of data parallelism. We show both the computation (even rows) and the data-parallel communication (odd rows), assumed to run in parallel CUDA streams. (We omit the pipeline-parallel communication for simplicity.) Looped schedules run significantly faster than their non-looped counterparts, with \lbf being the fastest.
    }
    \label{fig:pipe_schedules}
\end{figure*}

There are two common schedules for pipeline parallelism: with GPipe (\nlbf) \cite{gpipe}, the entire forward pass is run first, followed by the backward pass (Figure \ref{fig:schedule_gpipe}), while with 1F1B (\nldf) \cite{pipeDream}, the forward and backward steps are alternated so that earlier micro-batches finish as soon as possible. The two schedules have the same computational efficiency, but \nldf uses less activation memory.

Pipeline parallelism alone can in theory train moderately large models but is impractical as its scaling is limited by the depth of the model. Instead, \pp is most relevant when combined with \ddp because it may lower the gradient reduction overhead for a low batch size per GPU. For a fixed batch size, it divides all of \tnet, \tover and \tcomp,
so in terms of \bgpu the efficiency condition becomes
\begin{equation}
    \bgpu\geq\frac{\mbc\btmin}\npp \quad\text{or}\quad \bgpu\gg\frac\btmin\npp.
    \label{eq:pipeline_efficiency}
\end{equation}
While both equations appeared to be improved when compared to Eq. (\ref{eq:ddp_efficiency}), the overlapped equation is in general \emph{worse} due to the high number of sequential micro-batches required for \pp. On the other hand, the non-overlapped condition is less constraining, and with a high enough \npp the overhead may be minimal even at \bmin.

An important caveat when combining \ddp and \pp is that it excludes \fsdp. \pp requires gradient accumulation, so combining with \fsdp would require a repetition of the network operations, making the data-parallel network usage even worse than with \ddp alone. Instead, \nsdp or \psdp should be used, and a high \npp may be needed to limit the training state memory usage.

There has been recent progress in reducing the size of the pipeline bubble. \emph{Chimera} \cite{Li_2021} achieves it with a hybrid of data and pipeline parallelism where each device stores multiple pipeline stages, so that it is only idle when \emph{all} the stages are input-starved. However, Chimera requires additional memory and data-parallel network communication, which complicates its use for larger models. An alternative method, \emph{looping pipelines}, introduced in \cite{nvidiaMegatron2}, shrinks the bubble by storing multiple smaller, non-consecutive stages per device (Figure \ref{fig:ring_placement}). Looping pipeline avoid the memory and data-parallel network overhead of Chimera, though they require  extra pipeline-parallel communication. They are the discussed in more details in Section \ref{sec:breadth_first_pipeline}.

\subsection{Tensor parallelism}
\label{subsec:tensor}

Tensor parallelism (\tp) is another form of model parallelism, dividing the model along its \emph{width} \cite{googleMesh,nvidiaMegatron1}. By extension, it also divides the training state and reduces its memory usage. Each of the \ntp tensor-parallel devices processes a subset of the channels for the \emph{same} samples, and shares intermediate activations as needed. In particular, it has no requirement on the batch size, so $\bmin=\ntp^{-1}$.
However, the high network usage of \tp (which increases with \ntp) requires an extremely fast network such as NVLink, generally restricting \tp to a single node.

Although tensor parallelism scales poorly in isolation, it can be used in combination with other methods to reduce the memory usage and train with a lower batch size per GPU. Following the same reasoning as for Eq. (\ref{eq:pipeline_efficiency}), we find
\begin{equation}
    \bmin=\frac{1}{\ntp};
    \quad \bgpu
    \geq\frac{\mbc\btmin}{\npp\ntp}
    \quad\text{or}\quad
    \bgpu\gg\frac{\btmin}{\npp\ntp},
    \label{eq:pipeline_efficiency_3d}
\end{equation}
i.e., \tp divides both \bmin and the minimum \bgpu needed for \ddp network efficiency.

\subsection{State-of-the-art}
\label{subsec:combined_methods}

As \fsdp and \pp are mutually exclusive, there are two options for training large language models. 

The combination of all three base methods (\ddp, \pp and \tp), \textbf{3d parallelism}, was the first to successfully train large language models. This method scales well to large clusters, but generally has a lower GPU utilization due to the pipeline bubble and poor data-parallel network overlap. 3d parallelism was for example used to train GPT-3 (175 B parameters) \cite{openAiGpt3} and Megatron-Turing NLG (530 B, with \psdp) \cite{microsoftTuring}. Although we expect the looped, depth-first pipelines of \cite{nvidiaMegatron2} to be the most efficient version of 3d parallelism, we also treat non-looped pipelines as state-of-the-art, as they are still widely used and may be more efficient in certain cases (as demonstrated in Section \ref{sec:evaluation}).

Alternatively, the combination of \fsdp and \tp, (a form of) \textbf{2d parallelism}, generally allows for a higher GPU utilization, but does not scale as well due to the strict requirement on the batch size per GPU. 2d parallelism has been successfully used to train OPT (175 B) \cite{facebookOpt} and PaLM (540 B parameters) \cite{googlePalm}, which also used the advantageous network structure of the TPU pod to lower \bgpu. 

Breadth-First Pipeline Parallelism, introduced in the next section, offers a third option which can efficiently mix \fsdp with \pp. It also combines the low batch size per GPU of 3d parallelism with the computational efficiency of 2d parallelism.

\subsection{Effect of the batch size}
\label{subsec:critical_batch}

In stochastic gradient descent, a (mini-)batch is used to approximate the true gradients of the weights with respect to the loss. Increasing the batch size \bs generally improves this approximation, leading to more efficient steps. For small batches, this is computationally efficient, with larger batches allowing to train for proportionally fewer steps, for a near-constant computing power. However, for large batches the approximation is already accurate and additional samples provide a negligible improvement, leading to a waste of computing power. 


Empirically, the number of samples needed to reach a given validation loss has been shown to follow the curve \cite{criticalBatch}
\begin{equation}
    \text{Samples}\propto 1+\frac{\bs}{\bcrit},
    \label{eq:critical_batch}
\end{equation}
where the \emph{critical batch size} \bcrit depends on the model, training scheme and target validation loss,
and it can be estimated by measuring the gradient statistics (see Appendix \ref{sec:critical_batch_theory}). In short, the relative overhead is equal to the ratio $\bs/\bcrit$. For example, GPT-3 was trained with a batch size of 3 million tokens, with a critical batch size estimated to 10 million tokens \cite{openAiScaling}, for an overhead of about 30\%.  


For both state-of-the-art methods, the number of GPUs \ngpu can be scaled with minimal impact on the GPU utilization, provided the batch size per GPU \bgpu is kept constant. 
Therefore, Eq. (\ref{eq:time_cost}) can be rewritten as a trade-off with respect to \ngpu, assuming the cost is proportional to the number of samples processed (\ref{eq:critical_batch})
\begin{equation}
    \text{Cost}\propto 1+\bgpu\frac{\ngpu}{\bcrit},\qquad
    \text{Time}\propto \frac{\text{Cost}}{\ngpu}.
    \label{eq:time_cost_tradeoff}
\end{equation}



\section{Breadth-first pipeline}
\label{sec:breadth_first_pipeline}

\begin{figure}[t]
    \centering
    \subcaptionbox{
        Theoretical efficiency
        \label{fig:vary_batch_size_theory_overlapped}
    }{
        \includegraphics[scale=0.5]{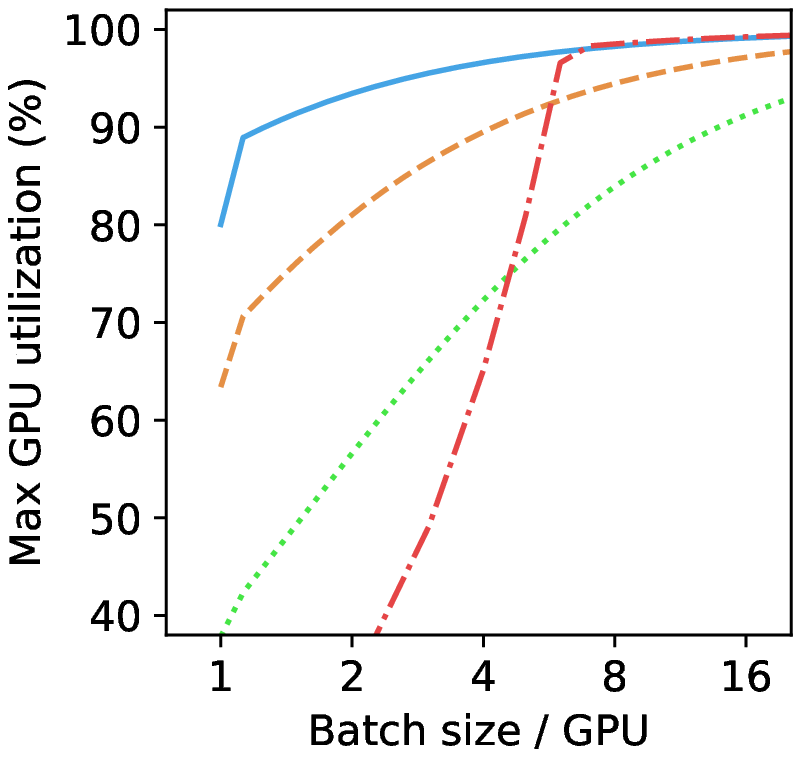}
    }\subcaptionbox{
        Without network overlap
        \label{fig:vary_batch_size_theory_no_overlap}
    }{
        \includegraphics[scale=0.5]{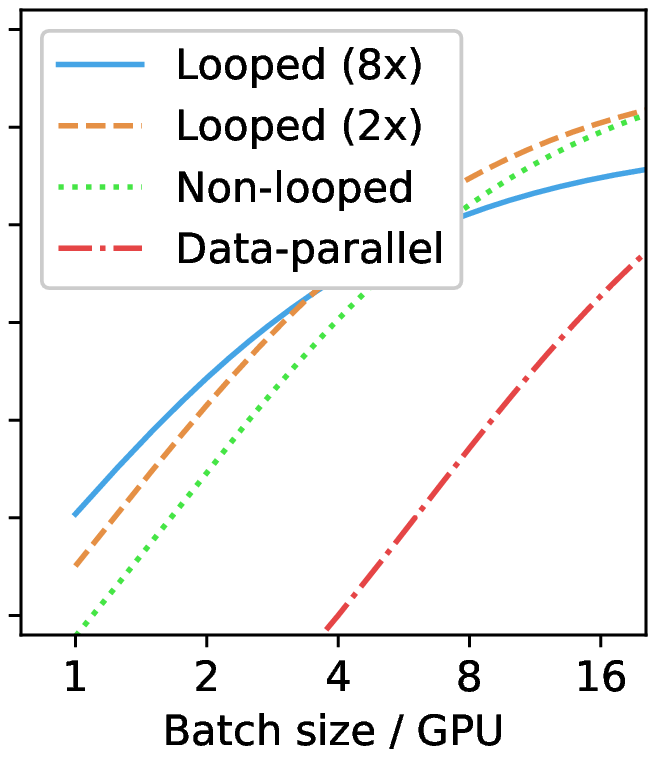}
    }
    \caption{(a) Comparison of the theoretical efficiency as a function of the batch size per GPU for looped and non-looped pipelines, and for pure data parallelism, for and example with $\btmin=6$, \ntp=1. Note the jump near $\bmin=1$ related to the pipeline-parallel network overlap.
    (b) The theoretical efficiency for the same configurations without data and pipeline network overlap, shown to emphasize the renewed importance of overlap for looped pipelines.
    }
    \label{fig:vary_batch_size_theory}
\end{figure}

\begin{figure}[tp]
    \centering
    \subcaptionbox{
        Standard
        \label{fig:line_placement}
    }{
        \includegraphics{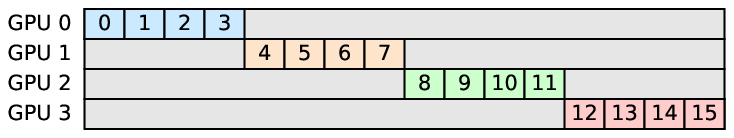}
    }\hspace{-8pt}\subcaptionbox{
        Looping
        \label{fig:ring_placement}
    }{
        \includegraphics{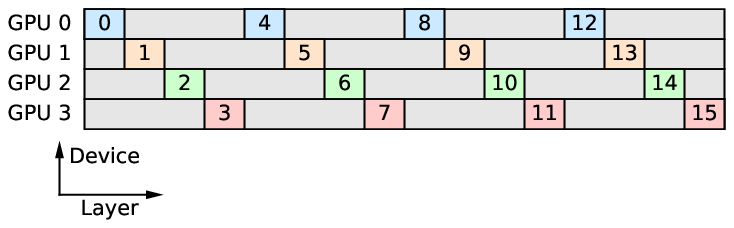}
        \vspace{-0.375in}
    }
    \vspace{0.05in}
\caption{Comparison of the standard and looping layer placements for a 16-layer model. The numbers (and x axis) show the index of the layers.}
\label{fig:pipeline_placement}
\end{figure}

In this section, we introduce our method, Breadth-first Pipeline Parallelism. We begin by introducing the two main components, looping pipelines and the breadth-first schedule, then present their benefits to large language model training from a theoretical perspective. We also briefly survey other use cases for our method.


\subsection{Looping pipeline and breadth-first schedule}

As described in section \ref{subsec:pipeline}, pipeline parallelism typically splits the layers into a single stage per device (Figure \ref{fig:line_placement}). This linear topology 
minimizes network communication but suffers heavily from the pipeline bubble. 
In a looping pipeline, first introduced in \cite{nvidiaMegatron2}, we instead divide the network into a large number of (identical or near-identical) stages (\nstage), wrapping them around by connecting the first and last device to form a ring (or more precisely a coil), looping $\nloop=\tfrac{\nstage}{\npp}$ times (Figure \ref{fig:ring_placement}). With this method, data reaches the last device after traversing a fraction of the layers, so the bubble overhead is reduced to 
\begin{equation}
    \text{Bubble}=\frac{\npp-1}{\mbc\nloop}.
    \label{eq:bubble_looped}
\end{equation}


In a looping pipeline, a given device can only process a single stage at once, even if there is queued input on multiple of them. The schedule may either prioritize earlier micro-batches (\emph{depth-first}), running micro-batches in ``sequences'' of \npp, or earlier stages (\emph{breadth-first}), running all micro-batches at once. These two options pair naturally with the backward-first approach of \nldf and the forward-first approach of \nlbf, respectively. We call the resulting methods \emph{depth-first pipeline} (\ldf) and \emph{breadth-first pipeline} (\lbf). The former, suggested in \cite{nvidiaMegatron2}, allows lowering the activation memory but only for a large number of micro-batches, i.e., in the scenario we are trying to avoid. It also constrains \mbc to a multiple of \npp. Instead, we argue that latter is preferable, 
allowing to train more efficiently with a low batch size per GPU.

\subsection{Analysis}
\label{subsec:bf_theory}

According to Equation (\ref{eq:bubble_looped}), computational efficiency now requires $\mbc\nloop\gg\npp$, so no longer strictly needs a high batch size. Instead, it can be achieved by maximizing \nloop, which is however far from trivial.
By its definition, a high \nloop requires a high \nstage and a small \npp.

Increasing \nstage is straightforward but adds extra pipeline-parallel network communication. The network usage remains small from a bandwidth perspective, but in practice this communication has a major impact on performance due to the small but numerous latency and synchronization overheads (see Section \ref{subsec:result_stage_size}). This overhead can be largely avoided by overlapping the transfers with computation. The depth-first schedule as introduced in \cite{nvidiaMegatron2} does not allow such overlap, since the transfers introduce delays in the pipeline which prevent the micro-batches from looping around when expected, causing the first device to be input-starved. (We believe (but did not verify) this can be addressed by running with sequences of more than \npp micro-batches, essentially forming a hybrid between the two schedules.) The breadth-first schedule, on the other hand, allows for such overlap when $\mbc>\npp$, because the $\mbc-\npp$ extra micro-batches can absorb the delay. The increase in batch size is unavoidable because micro-batches cannot take part in computation while being transferred, though a single extra micro-batch is generally sufficient.

As \nstage is limited to the number of layers in the model, increasing it alone may not be enough for a high \nloop. For example, in \cite{nvidiaMegatron2} a 128-layer, trillion-parameter model was trained with $\npp=64$, constraining to $\nloop\leq2$. Further progress can be made by reducing \npp, however such small pipelines go against the recommendations of Section \ref{subsec:pipeline}, which suggested a large \npp (1) to limit the memory usage of the training state and (2) to reduce the data-parallel network overhead. 

First, in a non-looping pipeline, a large \npp is needed to limit the training state memory usage for large models. \fsdp is inefficient as it involves a repetition of the weight reconstruction and gradient reduction for every micro-batch. A breadth-first schedule avoids any such repetition as it aggregates the steps by layer, so each layer is reconstructed for a single pair of contiguous intervals. The depth-first schedule also improves on non-looped pipelines, but requires a repetition for every micro-batch sequence, and has twice as many reconstructed layers when alternating between the forward and backward passes.

Second, according to Eq. (\ref{eq:pipeline_efficiency_3d}), a large \npp may be needed to minimize overhead from the data-parallel network operations, especially because these operations are poorly overlapped with computation. This requirement is often ignored in the literature because it's already avoided through a high batch size per GPU and/or large model parallelism. For example, \cite{nvidiaMegatron2} selects \npp according to memory usage only, which is justified as the factor $\bgpu\npp\ntp$ ranges from 48 to 512, so is always well above $\btmin\approx4$ (see Appendix \ref{subsubsec:network_data}). 
However, this is not necessarily the case for a low batch size per GPU, as Eq. (\ref{eq:pipeline_efficiency_3d}) reduces to $\npp\gg\btmin$ at \bmin. With a looping pipeline, the overlap is greatly improved: instead of a single micro-batch, \ldf overlaps with a sequence of \npp micro-batches, while \lbf overlaps with the entire batch. Thus, Breadth-First Pipeline Parallelism has the best network overlap, with a milder efficiency condition
\begin{equation}
    \bgpu\geq\frac\btmin{\npp\ntp},
    \label{eq:pipeline_efficiency_bf}
\end{equation}
which makes it more efficient for a low \npp and a low \bgpu. This condition also sets a lower bound on \npp for efficient training at \bmin (and an upper bound on \nloop), $\npp\geq\tfrac\btmin\bmin$.

\begin{table*}[tp]
{
\centering
\caption{Relative performance of distributed training methods on a large cluster ($\ndp\gg1$). For large models with a small batch size per GPU, the important quantities are the pipeline bubble, state memory and DP network overhead. Only Breadth-First Pipeline Parallelism achieves good performance on all three, with the depth-first being a close second when \fsdp is not required.}
\label{tab:method_comparison}
\tiny
\begin{tabular}{ccccccccccc}
\toprule
Method                & Pipeline         & State                     & Activation                   & DP                      & DP                           & PP          & Easy PP     & Flexible \\
                      & bubble           & memory$^1$                & memory$^2$                   & network                 & overlap                      & Network     & overlap$^3$ & \mbc   \\
\midrule
No pipeline           & 0  \tu\tu        & $\nl$ \td\td              & $\mbs$ \tu                   & $2$ \td\td              &$ (1-1/\nl)/\mbc$ \tu\tu$^4$   & 0 \tu\tu    & N.A.        & \tu\td$^4$ \\
No pipeline (\fsdp)   & 0  \tu\tu        & $2$ \tu\tu                & $\mbs$ \tu                   & $3\mbc$ \td\td\td$^4$   &$ (1-1/\nl)/\mbc$ \tu\tu$^4$  & 0 \tu\tu    & N.A.        & \tu\td$^4$     \\
GPipe                 & $1$ \td\td       & $\nl/\npp$ \tu            & $\mbs\mbc/\npp$ \tu\td       & $2/\npp$ \tu\tu         & $(1-\npp/\nl)/\mbc$ \td\td    & 1 \tu       & \tu         & \tu  \\
1F1B                  & $1$ \td\td       & $\nl/\npp$ \tu            & $\lessapprox2\mbs$ \tu       & $2/\npp$ \tu\tu         & $(1-\npp/\nl)/\mbc$ \td\td    & 1 \tu       & \td         & \tu    \\
1F1B  (\fsdp)         & $1$ \td\td       & $2$ \tu\tu                & $\lessapprox2\mbs$ \tu       & $3\mbc$ \td\td\td\td\td & $1-\npp/\nl$ \tu              & 1 \tu       & \td         & \tu    \\
Chimera$^5$           & $1/\chimera$ \tu & $\chimera \nl/\npp$ \td   & $\leq2\mbs$ \tu              & $2\chimera/\npp$ \td    & $\approx1-1/\chimera$ \tu\td  & 1 \tu       & \tu\td      & \td  \\
Depth-first           & $1/\nloop$ \tu   & $\nl/\npp$ \tu            & $\lessapprox\mbs+\mbs/\nloop$ \tu& $2/\npp$ \tu\tu     & $(1-\npp/\nl)\npp/\mbc$ \td   & \nloop \td  & \td         & \td  \\
Breadth-first         & $1/\nloop$ \tu   & $\nl/\npp$ \tu            & $\mbs\mbc/\npp$ \tu\td       & $2/\npp$ \tu\tu         & $1-\npp/\nl$ \tu              & \nloop \td  & \tu         & \tu  \\
Breadth-first (\fsdp) & $1/\nloop$ \tu   & $2$  \tu\tu               & $\mbs\mbc/\npp$ \tu\td       & $3/\npp$ \tu            & $1-\npp/\nl$ \tu              & \nloop \td  & \tu         & \tu   \\
\bottomrule
\end{tabular}
\vspace{10pt}

}
\scriptsize{
$^1$ Assuming \psdp (or \fsdp), otherwise the values are multiplied by 3 or more. (Appendix \ref{subsubsec:memory_state})\\
$^2$ These values are somewhat misleading, as the non-pipelined methods typically have a higher micro-batch size so generally need \emph{more} activation memory.
At \btmin, all values are equal. With activation checkpointing, this represents the checkpoint memory, and the layer activations and gradients have the same memory usage for all methods.\\
$^3$ We believe PP overlap is feasible to some extent for all schedules, but 1F1B and Depth-first add significant complications and need to be modified to support it, while Chimera has only been shown to support it for $\mbc\geq2\npp$.\\
$^4$ The qualitative performance of the non-pipeline approach assumes $\mbc=1$, otherwise it is much worse. In Appendix \ref{sec:breadth_first_2d}, we present a breadth-first gradient accumulation method which allows for a good data-parallel performance with $\mbc>1$ at the cost of extra activation memory.\\
$^5$ Chimera with \chimera pipelines, where \chimera is an even integer (typically 2). We assume `forward doubling` and `backward halving` (\cite{Li_2021}, Section 3.5), which reduces the pipeline bubble and allows for PP overlap for $\mbc\geq2\npp$ at the cost of extra activation memory.
}

\end{table*}

In summary, a breadth-first schedule trains more efficiently with a high \nloop and thus for a low batch size per GPU, because it allows for better overlap of the data and pipeline-parallel network communication (Figure \ref{fig:vary_batch_size_theory}), and combines better with \fsdp. Additional details can be found in Table \ref{tab:method_comparison}, as well as a comparison with alternative methods.

One caveat of our analysis is that the two schedules are relatively similar at the minimum batch size per GPU \bmin, which is precisely the value we would like to use. At \bmin, both fully overlap the data-parallel operations, and the depth-first schedule is only slightly less effective with \fsdp. For a slightly higher batch size, both should allow for pipeline-parallel network overlap (though we only verified this for \lbf). However, this similarity disappears when considering realistic, non-ideal scenarios. While a looping pipeline significantly reduces the impact of the batch size per GPU, it does not eliminate it. There are still benefits to training with a larger batch size, for example because \nloop can only be increased up to a certain point (Figure \ref{fig:vary_batch_size_theory}). The exact batch should be selected such that it minimizes the training cost and time when taking into account both the GPU utilization and the batch size overhead (\ref{eq:critical_batch}). For larger clusters, the optimal batch size is near \bmin, but smaller ones offer more flexibility.

\subsection{Additional use cases}

Although Breadth-First Pipeline Parallelism is aimed at training large language models as described in section \ref{sec:extended_intro}, it is also useful in other scenarios. Most importantly, the improved network overlap makes the method well suited for slower networks, for example on GPU clusters without InfiniBand support, that are instead only connected through a slow Ethernet network. This is the for example the case for many cloud platforms and older clusters. In that case, it is more difficult to minimize the data-parallel network overhead, and Breadth-First Pipeline Parallelism is expected to perform more efficiently and at a lower batch size per GPU due to its advantageous network efficiency condition (\ref{eq:pipeline_efficiency_bf}). The justification for this is identical to the analysis of Section \ref{subsec:bf_theory}, with a high \btmin instead of a low \npp. Note that this overlap is achieved with looping which affects the condition (\ref{eq:pipeline_efficiency_bf}) (smaller \npp), so training is still not optimal and is expected to require a batch size above \bmin.

\section{Evaluation}
\label{sec:evaluation}

\begin{table}[tp]
    \caption{Details of the models}
    \centering\footnotesize
    \begin{tabular}{|c|ccccc|}
        \toprule
        Model & Num & Attention & Head & Hidden & Seq \\
         & layers & heads & size & size & length \\
        \midrule
        52 B & 64 & 64 & 128 & 8192 & 1024 \\
        6.6 B & 32 & 32 & 128 & 4096 & 1024 \\
        \bottomrule
    \end{tabular}
    \label{tab:model_descriptions}
\end{table}

We ran a series of experiments to verify our main claim, namely that breadth-first pipeline parallelism allows for a faster and/or cheaper training of large language models, under the assumptions described in section \ref{sec:extended_intro}.

We ran our experiments on a cluster of eight DGX-1 nodes, for a total of 64 V100-SXM2-32GB GPUs, connected through an InfiniBand network. All our experiments were run on the same hardware, except for one node which was temporarily replaced due to a hardware failure. Although we tried our best to also use the same software, our implementation (described in Appendix \ref{sec:implementation}) does not support the 1F1B and depth-first pipeline schedules, for which we used Megatron-LM \cite{nvidiaMegatron2} instead. As Megatron-LM does not support (data and pipeline-parallel) network overlap or \psdp, our results may somewhat underestimate the performance for these schedules.\footnote{Megatron-LM added support for PSDP (``distributed optimizer'') in a later version, published alongside \cite{nvidiaMegatron3}, but it could not be included in our experiments which were already underway.} We tested two different models (with a BERT architecture): a moderately large, 52 billion-parameter model, and a smaller, 6.6 billion-parameter model (Table \ref{tab:model_descriptions}).

\subsection{Fixed configurations}
\label{subsec:result_fixed_config}

\begin{figure}[t]
    \centering
    \subcaptionbox{
        52 B model
        \label{fig:bert_52b_mb_sizes}
    }{
        \includegraphics[scale=0.5]{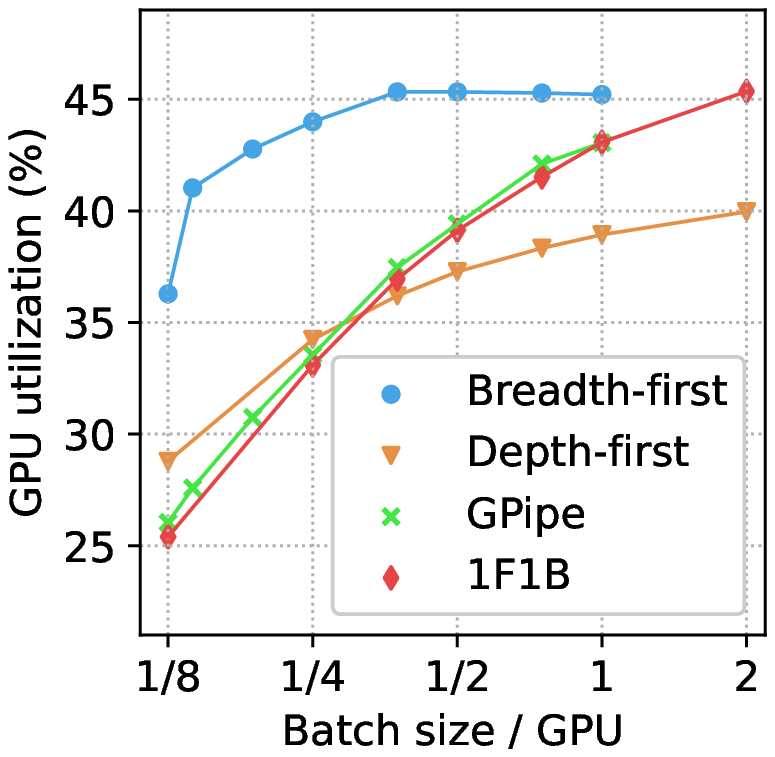}
    }\subcaptionbox{
        6.6 B model
        \label{fig:bert_6607m_mb_sizes}
    }{
        \includegraphics[scale=0.5]{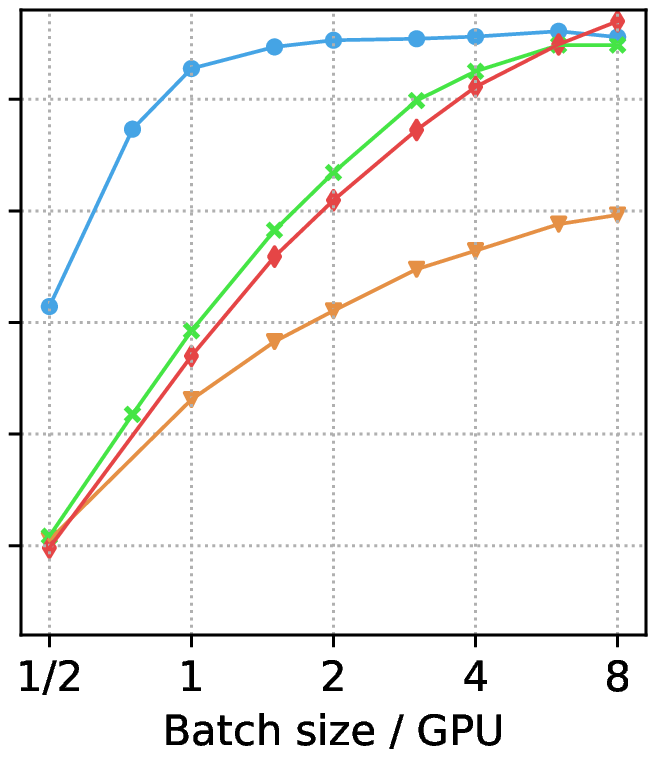}
    }
    \caption{GPU utilization for the looped ($\nloop=4$) and non-looped schedules as a function of the number of sequential micro-batches ($\mbs=1$) for (a) the 52 B model ($\npp=\ntp=8$, $\ndp=1$) and (b) the 6.6 B model ($\npp=4$, $\ntp=2$, $\ndp=8$).
    }
    \label{fig:bert_mb_sizes}
\end{figure}

We first compared the four pipeline schedules for matching configurations. In figure \ref{fig:bert_mb_sizes}, we show the GPU utilization as a function of the batch size per GPU (with a fixed micro-batch size) for both models, with fixed distributed configurations. The results are qualitatively similar to the theoretical predictions of Figure \ref{fig:vary_batch_size_theory}, where the depth-first and 1F1B schedules are not overlapped (Figure \ref{fig:vary_batch_size_theory_no_overlap}).\footnote{We recall that GPipe and 1F1B should have exactly the same performance, other than GPipe running out of memory for larger batch sizes. Thus, the observed difference can be entirely attributed to implementation differences, in particular the lack of network overlap in Megatron-LM (1F1B).} For smaller batches, the breadth-first schedule is by far the most efficient, minimizing both the bubble and network overheads. The depth-first schedule also reduces the pipeline bubble, but its high network overhead makes the performance \emph{worse} than than the non-looped configurations in most cases. For larger batches, the pipeline bubble is small in all cases, and 1F1B is the fastest because of its lower pipeline-parallel network overhead and memory usage. Note that while Figure \ref{fig:bert_mb_sizes} supports our claims, it does not reflect the full potential of each method. A more unbiased comparison is performed in section \ref{subsec:result_batch_size}.

\subsection{Bubble and network overheads}
\label{subsec:result_stage_size}

\begin{figure}[t]
    \centering
    \subcaptionbox{
        $\bs=16$
        \label{fig:bert_52b_stage_small_batch}
    }{
        \includegraphics[scale=0.5]{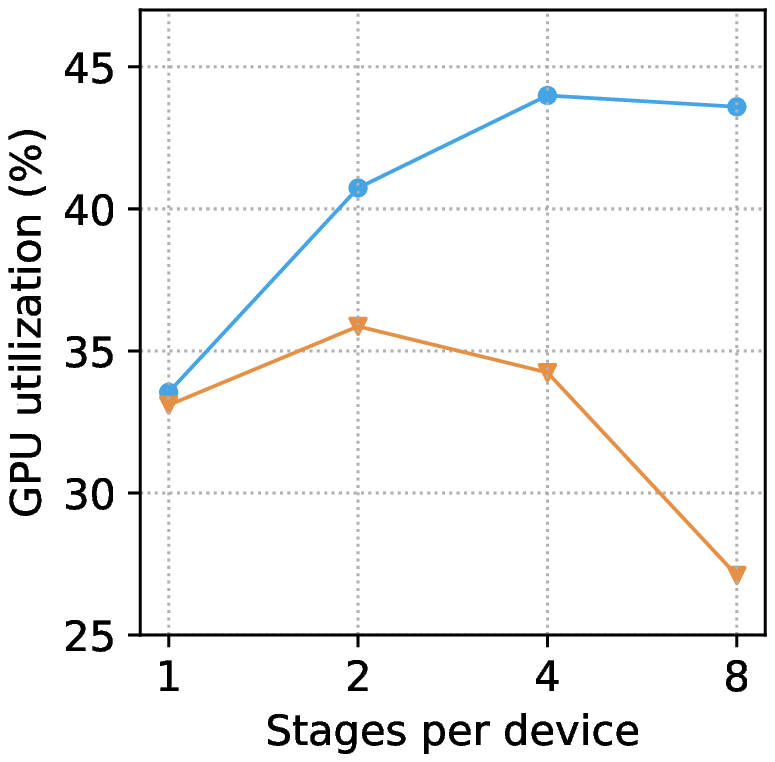}
    }\subcaptionbox{
        $\bs=64$
        \label{fig:bert_52b_stage_large_batch}
    }{
        \includegraphics[scale=0.5]{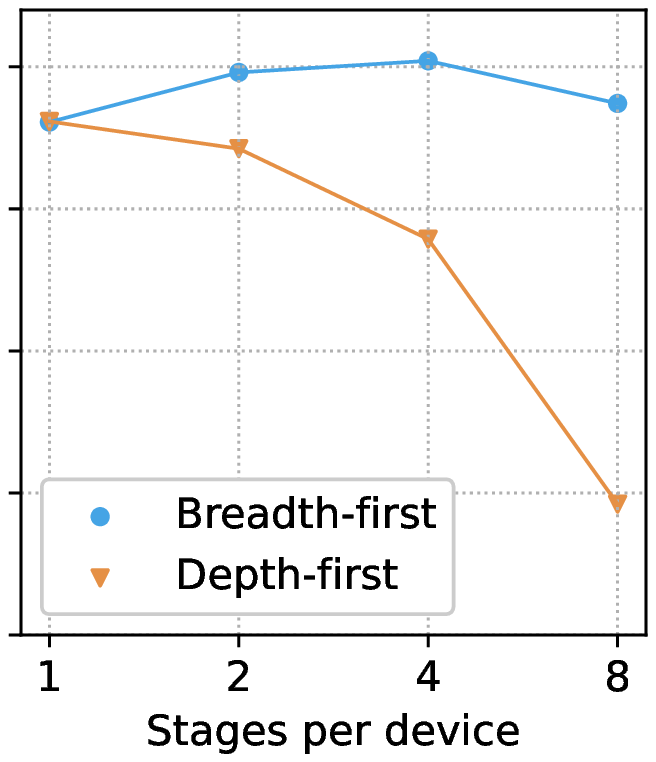}
    }
    \caption{GPU utilization for the breadth-first (ours) depth-first (Megatron-LM) schedules as a function of the number of stages per device \nloop for the 52 B model for two different batch sizes ($\npp=\ntp=8$, $\ndp=1$, $\mbs=1$).
    }
    \label{fig:bert_52b_stage}
\end{figure}

To quantify the relative impact of the bubble and pipeline-parallel network overhead, we compared the breadth-first and depth-first schedules as a function of \nloop (including $\nloop=1$, which corresponds to GPipe and 1F1B, respectively). The results are shown in Figure \ref{fig:bert_52b_stage} for the 52 B model with two different batch sizes. Both schedules benefit from the bubble reduction, especially at a smaller batch size (Figure \ref{fig:bert_52b_stage_small_batch}), but the network overhead is far more important for the depth-first schedule, which only benefits from looping for small batch sizes and small \nloop (as already observed in Figure \ref{fig:bert_mb_sizes}). From Figure \ref{fig:bert_52b_stage_large_batch}, we estimate that the network overhead is at least 40\% for $\nloop=8$ (30\% utilization vs 43\% for $\nloop=1$). This is far higher than the value of $1.6\%$ predicted from the bandwidth usage (Appendix \ref{subsubsec:network_pipe}), which suggests the overhead is mainly due to latency and synchronization, as claimed in Section \ref{subsec:bf_theory}. The breadth-first schedule avoids most but not all of this overhead with network overlap, resulting in $\nloop=4$ being the optimal value for the present case.

\begin{figure*}[tp]
    \centering
    \subcaptionbox{
        52 B model
        \label{fig:bert_52b_vary_batch_size}
    }{
        \includegraphics[scale=0.72]{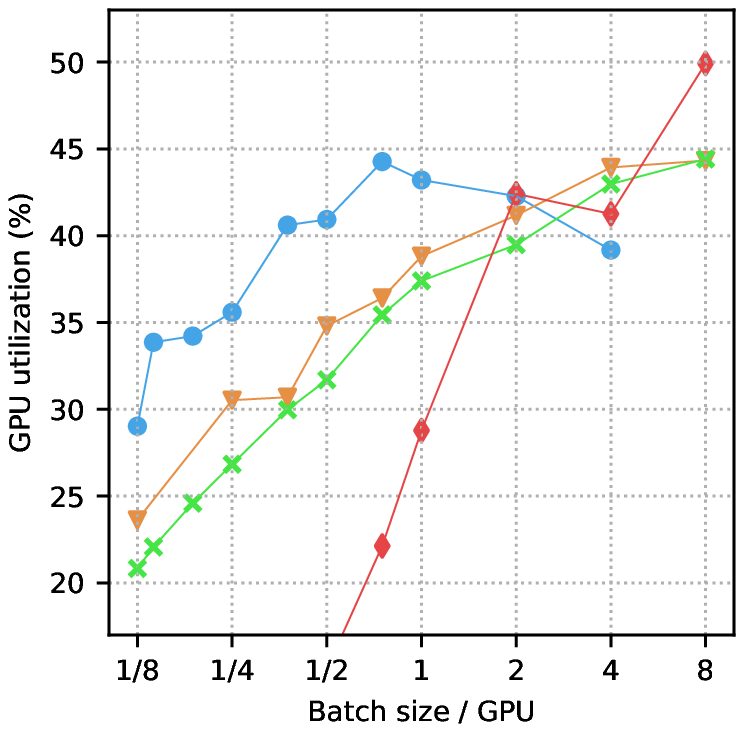}
    }
    \hspace{-10pt}\subcaptionbox{
        6.6 B model
        \label{fig:bert_6607m_vary_batch_size}
    }{
        \includegraphics[scale=0.72]{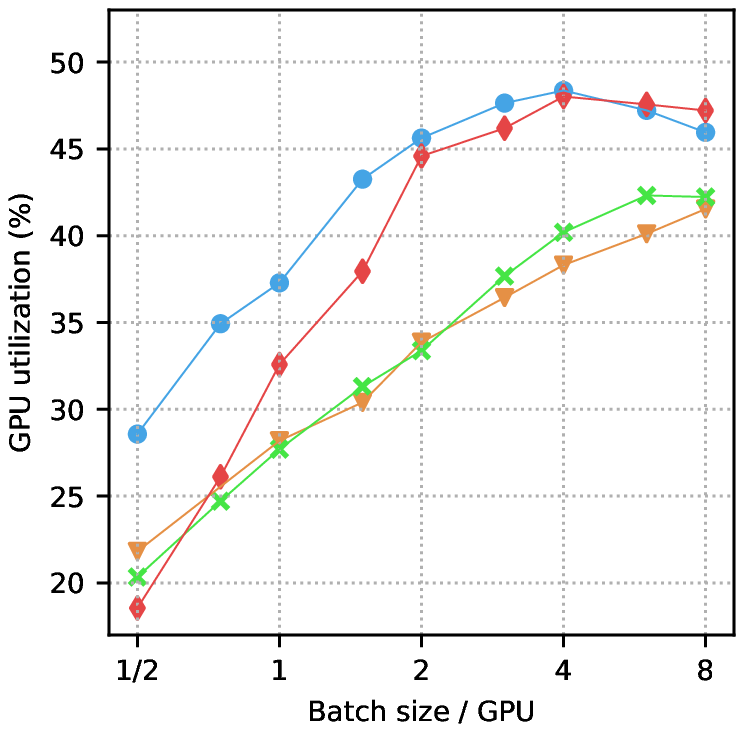}
    }
    \hspace{-10pt}\subcaptionbox{
        6.6 B model, Ethernet
        \label{fig:bert_6607m_eth_vary_batch_size}
    }{
        \includegraphics[scale=0.72]{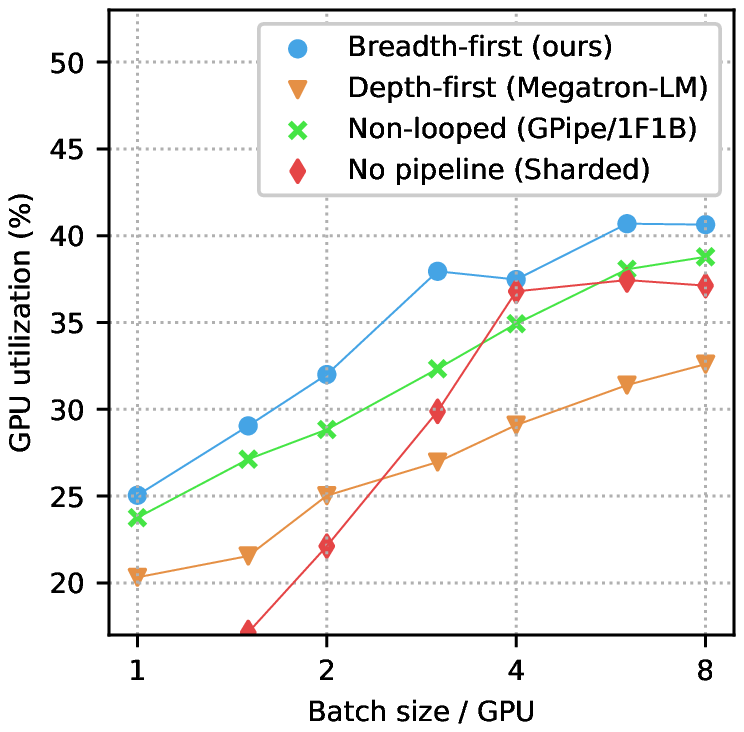}
    }
    \caption{Highest GPU utilization observed on a cluster of 64 V100 GPUs for the selected methods, as a function of the batch size.
    }
    \label{fig:bert_vary_batch_size}
\end{figure*}

\subsection{Optimal configurations}
\label{subsec:result_batch_size}

While the results of Section \ref{subsec:result_fixed_config} agree with our theoretical prediction, they use sub-optimal configurations and thus underestimate the true performance of the methods. For example, the breadth-first schedule stands to benefit from \fsdp and reduced model parallelism at higher batch sizes, while Megatron-LM is more efficient with a higher micro-batch size and fewer stages per device. The fixed configuration also prevented a comparison with a non-pipelined approach which evidently does not admit a configuration with $\npp>1$.

To achieve a fair comparison, we performed an extensive search over the configuration spaces and selected the most efficient one for each method and for a selection of batch sizes. We compared Breadth-First Pipeline Parallelism against the three state-of-the-art methods introduced in Section \ref{subsec:combined_methods}: a depth-first pipeline as in \cite{nvidiaMegatron2}, a non-looped pipeline (GPipe or 1F1B), and no pipeline at all. In all cases, we also allowed for data and tensor parallelism. We performed the experiment for both the 52 B and 6.6 B models. To test the effect of our method for slow networks, we repeated the experiment for the smaller model a second time, where we disabled InfiniBand and instead trained using an Ethernet network. The results are shown in Figure \ref{fig:bert_vary_batch_size}.\footnote{The curves of Figure \ref{fig:bert_vary_batch_size} are not particularly smooth, which is a side effect of the optimization over the discrete configuration space. In other words, they result from a combination of many smooth curves such as those of Figure \ref{fig:bert_mb_sizes}, most of which do not cover the entire range of batch sizes, and jumps may occur at the edge of a curve.} See Appendix \ref{sec:detailed_results} for additional details, including a description of each optimal configuration and its memory usage.


For the larger, 52 B model (Figure \ref{fig:bert_52b_vary_batch_size}), the results roughly match the theory (Figure \ref{fig:vary_batch_size_theory}), and  breadth-first approach is the fastest at all but the largest batch size. Our method outperforms all other methods near $\beta_\text{min}=\tfrac{1}{8}$ (with one extra sample to allow for pipeline-parallel network overlap), running 53\% and 43\% faster than the non-looped and depth-first baselines, respectively. However, our method does benefit from larger batches, in large part by lowering the amount of tensor model parallelism (see Appendix \ref{sec:detailed_results}), which has a high overhead even for this model size. The non-pipelined approach does achieves higher utilization than our method, but for an excessively high batch size per GPU $\bgpu=8$ (even though the actual batch size has a perfectly reasonable value of 512, as will be demonstrated in section \ref{subsec:result_trade_off}). In this case, efficiency (nearly) plateaus at $\bgpu=2$, $\ntp=2$, suggesting $\btmin\approx4$.

For the smaller, 6.6 B model (Figure \ref{fig:bert_6607m_vary_batch_size}), our approach is again the most efficient, but by a smaller margin as the non-pipelined approach performs nearly as well (also with $\btmin\approx4$). This is largely because the model is at the lower end of ``large'' models, and our large model assumptions do not hold as well. For example, the micro-batch size has a noticeable influence on thread-level parallelism, and there is a high model-parallel overhead. None of the approaches is efficient at a low \bgpu.

The 6.6 B model can also be trained with an Ethernet network (Figure \ref{fig:bert_6607m_vary_batch_size}), though at a reduced efficiency. In that case, our method shows improvements for all \bgpu, and the non-pipelined approach is not as efficient even for a high \bgpu ($\btmin\approx 32$). The poor performance of Megatron-LM is largely attributed to the lack of network overlap.

\subsection{Trade-off}
\label{subsec:result_trade_off}

\begin{figure*}[tp]
    \centering
    \subcaptionbox{
        52 B model ($\bcrit\approx6780$)
        \label{fig:bert_52b_vary_cluster_size}
    }{
        \includegraphics[scale=0.72]{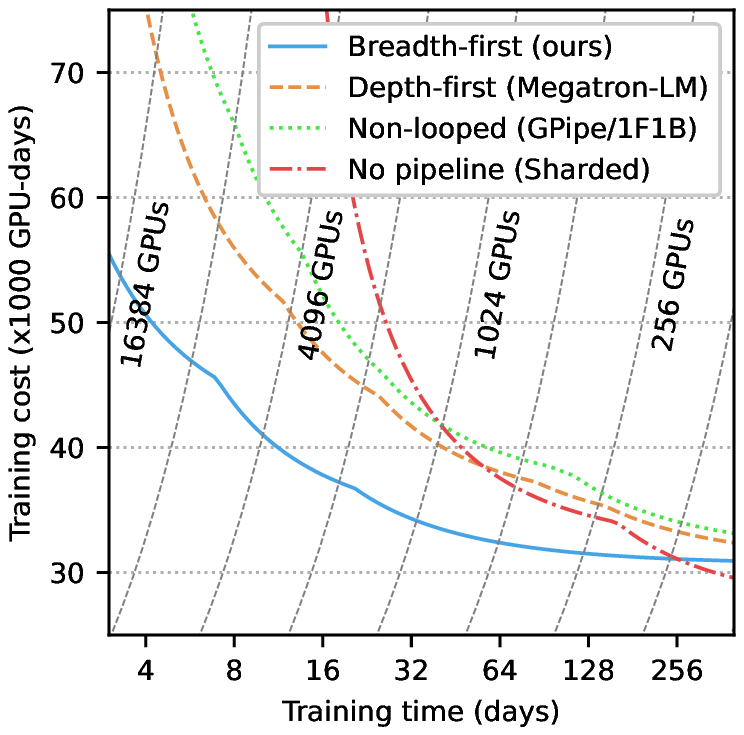}
    }
    \hspace{-10pt}\subcaptionbox{
        6.6 B model ($\bcrit\approx3430$)
        \label{fig:bert_6607m_vary_cluster_size}
    }{
        \includegraphics[scale=0.72]{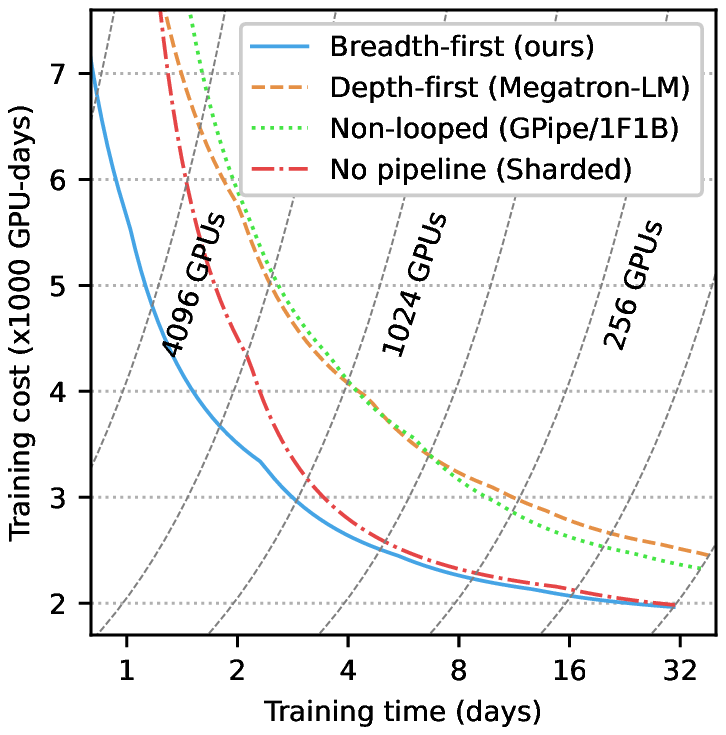}
    }
    \hspace{-10pt}\subcaptionbox{
        6.6 B model, Ethernet
        \label{fig:bert_6607m_eth_vary_cluster_size}
    }{
        \includegraphics[scale=0.72]{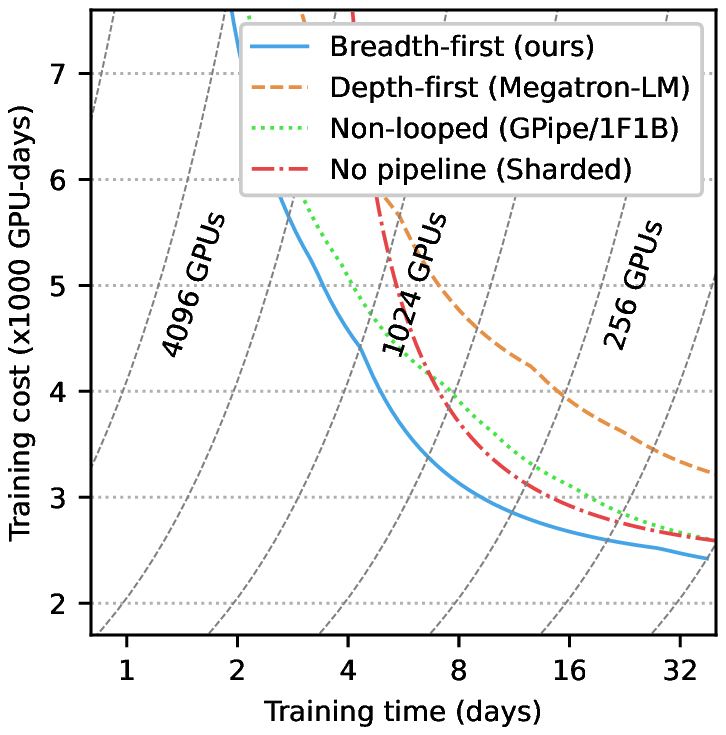}
    }
    \caption{Predicted trade-offs between the training cost and time, extrapolated from the results of Figure \ref{fig:bert_vary_batch_size}.}
    \label{fig:bert_vary_cluster_size}
\end{figure*}

Evaluating the training time and cost required some extrapolation as we did not have access to a sufficiently large cluster, and it was not possible to run training to completion. We extrapolated the results of Section \ref{subsec:result_batch_size} to a range of cluster sizes by scaling data parallelism with a constant batch size per GPU, assuming a constant GPU utilization. This is justified because it has almost no effect on the compute and network usages per GPU. We evaluated the training time for each extrapolation for a base training length of 50,000 times the critical batch size (347 and 176 billion tokens for the 52 B and 6.6 B model, respectively), scaled according to the high batch size overhead predicted from Eq. \ref{eq:critical_batch}. We used the results of \cite{openAiScaling} to estimate the critical batch size.
\footnote{These results were estimated by extrapolating the results for smaller models under a variety of assumptions, so come with a high uncertainty. Because of this (as well as other uncertainties, for example from the extrapolation and the batch size overhead \ref{eq:critical_batch} being approximate), our projected trade-offs in Figure \ref{fig:bert_vary_cluster_size} should not be considered as quantitatively exact. Note that the vast majority of the uncertainty only affects the scaling of the training time (x-axis).}\footnote{While the estimated critical batch sizes are higher than typical batch sizes, we remind that \ref{eq:critical_batch} still predicts a significant overhead. For example, a batch size of 1024 leads to an overhead of 15\% (52 B) or 30\% (6.6 B).}

For both models and each method, we selected the best extrapolation as a function of the cluster size, and plotted the associated times and costs (Figure \ref{fig:bert_vary_cluster_size}).
For the 52 B model, the breadth-first pipeline shows significant cost and time improvements at nearly all scales. The largest efficiency seen for the 2d approach is only meaningful for unreasonable training times of above six months. For the smaller model, our method shows improvements for all cluster sizes, though only significantly for larger clusters and at some extra cost.

\section{Conclusion}
\label{sec:conclusion}

We demonstrated that Breadth-First Pipeline Parallelism reduces the training time and cost of training large language models by combining a high efficiency with a low batch size per GPU. It minimizes both the network overhead and pipeline bubble, while its compatibility with \fsdp allows for more efficient configurations which would otherwise not be possible from a memory perspective. As a bonus, \fsdp should enable the training of much larger models than previously possible with pipeline-parallelism, with tens if not hundreds of trillions of parameters. This was already possible with 2D methods such as ZeRO-infinity \cite{microsoftZeroInfinity}, but with a prohibitively high training time. Although our method improves on that level, models of these sizes remain fundamentally limited in terms of \emph{both} training time and cost.

In the next step, we would like to evaluate our method on bigger models and with more modern hardware such as NVIDIA A100 or the upcoming H100. For example, Megatron-LM achieves its highest GPU utilization (57\%) for a 530 billion-parameter model on 280 A100s. We would also like to combine our method with other recent progress such as \emph{FlashAttention} \cite{flashAttention}, \emph{sequence parallelism} and \emph{selective activation recomputation} \cite{nvidiaMegatron3}. These methods are orthogonal to ours, so could be used to further improve training efficiency.

\section*{Acknowledgements}

The author is thankful to Harm de Vries for providing extensive support in writing the paper, and to Deepak Narayanan, Stefania Raimondo, Adam Salvail and Chris Tyler for reviewing and providing feedback.

\bibliographystyle{mlsys2023}
\bibliography{main}

\begin{thebibliography}{24}
\providecommand{\natexlab}[1]{#1}
\providecommand{\url}[1]{\texttt{#1}}
\expandafter\ifx\csname urlstyle\endcsname\relax
  \providecommand{\doi}[1]{doi: #1}\else
  \providecommand{\doi}{doi: \begingroup \urlstyle{rm}\Url}\fi

\bibitem[Brown et~al.(2020)Brown, Mann, Ryder, Subbiah, Kaplan, Dhariwal,
  Neelakantan, Shyam, Sastry, Askell, Agarwal, Herbert-Voss, Krueger, Henighan,
  Child, Ramesh, Ziegler, Wu, Winter, Hesse, Chen, Sigler, Litwin, Gray, Chess,
  Clark, Berner, McCandlish, Radford, Sutskever, and Amodei]{openAiGpt3}
Brown, T.~B., Mann, B., Ryder, N., Subbiah, M., Kaplan, J., Dhariwal, P.,
  Neelakantan, A., Shyam, P., Sastry, G., Askell, A., Agarwal, S.,
  Herbert-Voss, A., Krueger, G., Henighan, T., Child, R., Ramesh, A., Ziegler,
  D.~M., Wu, J., Winter, C., Hesse, C., Chen, M., Sigler, E., Litwin, M., Gray,
  S., Chess, B., Clark, J., Berner, C., McCandlish, S., Radford, A., Sutskever,
  I., and Amodei, D.
\newblock Language models are few-shot learners, 2020.
\newblock URL \url{https://arxiv.org/abs/2005.14165}.

\bibitem[Chowdhery et~al.(2022)Chowdhery, Narang, Devlin, Bosma, Mishra,
  Roberts, Barham, Chung, Sutton, Gehrmann, Schuh, Shi, Tsvyashchenko, Maynez,
  Rao, Barnes, Tay, Shazeer, Prabhakaran, Reif, Du, Hutchinson, Pope, Bradbury,
  Austin, Isard, Gur-Ari, Yin, Duke, Levskaya, Ghemawat, Dev, Michalewski,
  Garcia, Misra, Robinson, Fedus, Zhou, Ippolito, Luan, Lim, Zoph, Spiridonov,
  Sepassi, Dohan, Agrawal, Omernick, Dai, Pillai, Pellat, Lewkowycz, Moreira,
  Child, Polozov, Lee, Zhou, Wang, Saeta, Diaz, Firat, Catasta, Wei,
  Meier-Hellstern, Eck, Dean, Petrov, and Fiedel]{googlePalm}
Chowdhery, A., Narang, S., Devlin, J., Bosma, M., Mishra, G., Roberts, A.,
  Barham, P., Chung, H.~W., Sutton, C., Gehrmann, S., Schuh, P., Shi, K.,
  Tsvyashchenko, S., Maynez, J., Rao, A., Barnes, P., Tay, Y., Shazeer, N.,
  Prabhakaran, V., Reif, E., Du, N., Hutchinson, B., Pope, R., Bradbury, J.,
  Austin, J., Isard, M., Gur-Ari, G., Yin, P., Duke, T., Levskaya, A.,
  Ghemawat, S., Dev, S., Michalewski, H., Garcia, X., Misra, V., Robinson, K.,
  Fedus, L., Zhou, D., Ippolito, D., Luan, D., Lim, H., Zoph, B., Spiridonov,
  A., Sepassi, R., Dohan, D., Agrawal, S., Omernick, M., Dai, A.~M., Pillai,
  T.~S., Pellat, M., Lewkowycz, A., Moreira, E., Child, R., Polozov, O., Lee,
  K., Zhou, Z., Wang, X., Saeta, B., Diaz, M., Firat, O., Catasta, M., Wei, J.,
  Meier-Hellstern, K., Eck, D., Dean, J., Petrov, S., and Fiedel, N.
\newblock Palm: Scaling language modeling with pathways, 2022.
\newblock URL \url{https://arxiv.org/abs/2204.02311}.

\bibitem[Dao et~al.(2022)Dao, Fu, Ermon, Rudra, and Ré]{flashAttention}
Dao, T., Fu, D.~Y., Ermon, S., Rudra, A., and Ré, C.
\newblock Flashattention: Fast and memory-efficient exact attention with
  io-awareness, 2022.
\newblock URL \url{https://arxiv.org/abs/2205.14135}.

\bibitem[Devlin et~al.(2018)Devlin, Chang, Lee, and Toutanova]{bert}
Devlin, J., Chang, M.-W., Lee, K., and Toutanova, K.
\newblock Bert: Pre-training of deep bidirectional transformers for language
  understanding, 2018.
\newblock URL \url{https://arxiv.org/abs/1810.04805}.

\bibitem[Fedus et~al.(2021)Fedus, Zoph, and Shazeer]{googleSwitch}
Fedus, W., Zoph, B., and Shazeer, N.
\newblock Switch transformers: Scaling to trillion parameter models with simple
  and efficient sparsity, 2021.
\newblock URL \url{https://arxiv.org/abs/2101.03961}.

\bibitem[Goyal et~al.(2017)Goyal, Doll{\'a}r, Girshick, Noordhuis, Wesolowski,
  Kyrola, Tulloch, Jia, and He]{Goyal2017AccurateLM}
Goyal, P., Doll{\'a}r, P., Girshick, R.~B., Noordhuis, P., Wesolowski, L.,
  Kyrola, A., Tulloch, A., Jia, Y., and He, K.
\newblock Accurate, large minibatch sgd: Training imagenet in 1 hour.
\newblock \emph{ArXiv}, abs/1706.02677, 2017.

\bibitem[Harlap et~al.(2018)Harlap, Narayanan, Phanishayee, Seshadri, Devanur,
  Ganger, and Gibbons]{pipeDream}
Harlap, A., Narayanan, D., Phanishayee, A., Seshadri, V., Devanur, N., Ganger,
  G., and Gibbons, P.
\newblock Pipedream: Fast and efficient pipeline parallel dnn training, 2018.
\newblock URL \url{https://arxiv.org/abs/1806.03377}.

\bibitem[Hoffmann et~al.(2022)Hoffmann, Borgeaud, Mensch, Buchatskaya, Cai,
  Rutherford, Casas, Hendricks, Welbl, Clark, Hennigan, Noland, Millican,
  Driessche, Damoc, Guy, Osindero, Simonyan, Elsen, Rae, Vinyals, and
  Sifre]{googleChinchilla}
Hoffmann, J., Borgeaud, S., Mensch, A., Buchatskaya, E., Cai, T., Rutherford,
  E., Casas, D. d.~L., Hendricks, L.~A., Welbl, J., Clark, A., Hennigan, T.,
  Noland, E., Millican, K., Driessche, G. v.~d., Damoc, B., Guy, A., Osindero,
  S., Simonyan, K., Elsen, E., Rae, J.~W., Vinyals, O., and Sifre, L.
\newblock Training compute-optimal large language models, 2022.
\newblock URL \url{https://arxiv.org/abs/2203.15556}.

\bibitem[Huang et~al.(2018)Huang, Cheng, Bapna, Firat, Chen, Chen, Lee, Ngiam,
  Le, Wu, and Chen]{gpipe}
Huang, Y., Cheng, Y., Bapna, A., Firat, O., Chen, M.~X., Chen, D., Lee, H.,
  Ngiam, J., Le, Q.~V., Wu, Y., and Chen, Z.
\newblock Gpipe: Efficient training of giant neural networks using pipeline
  parallelism, 2018.
\newblock URL \url{https://arxiv.org/abs/1811.06965}.

\bibitem[Kaplan et~al.(2020)Kaplan, McCandlish, Henighan, Brown, Chess, Child,
  Gray, Radford, Wu, and Amodei]{openAiScaling}
Kaplan, J., McCandlish, S., Henighan, T., Brown, T.~B., Chess, B., Child, R.,
  Gray, S., Radford, A., Wu, J., and Amodei, D.
\newblock Scaling laws for neural language models, 2020.
\newblock URL \url{https://arxiv.org/abs/2001.08361}.

\bibitem[Korthikanti et~al.(2022)Korthikanti, Casper, Lym, McAfee, Andersch,
  Shoeybi, and Catanzaro]{nvidiaMegatron3}
Korthikanti, V., Casper, J., Lym, S., McAfee, L., Andersch, M., Shoeybi, M.,
  and Catanzaro, B.
\newblock Reducing activation recomputation in large transformer models, 2022.
\newblock URL \url{https://arxiv.org/abs/2205.05198}.

\bibitem[Li \& Hoefler(2021)Li and Hoefler]{Li_2021}
Li, S. and Hoefler, T.
\newblock Chimera.
\newblock In \emph{Proceedings of the International Conference for High
  Performance Computing, Networking, Storage and Analysis}. {ACM}, nov 2021.
\newblock \doi{10.1145/3458817.3476145}.
\newblock URL \url{https://doi.org/10.1145\%2F3458817.3476145}.

\bibitem[McCandlish et~al.(2018)McCandlish, Kaplan, Amodei, and
  Team]{criticalBatch}
McCandlish, S., Kaplan, J., Amodei, D., and Team, O.~D.
\newblock An empirical model of large-batch training, 2018.
\newblock URL \url{https://arxiv.org/abs/1812.06162}.

\bibitem[Narayanan et~al.(2021)Narayanan, Shoeybi, Casper, LeGresley, Patwary,
  Korthikanti, Vainbrand, Kashinkunti, Bernauer, Catanzaro, Phanishayee, and
  Zaharia]{nvidiaMegatron2}
Narayanan, D., Shoeybi, M., Casper, J., LeGresley, P., Patwary, M.,
  Korthikanti, V.~A., Vainbrand, D., Kashinkunti, P., Bernauer, J., Catanzaro,
  B., Phanishayee, A., and Zaharia, M.
\newblock Efficient large-scale language model training on gpu clusters using
  megatron-lm, 2021.
\newblock URL \url{https://arxiv.org/abs/2104.04473}.

\bibitem[Radford et~al.(2019)Radford, Wu, Child, Luan, Amodei, and
  Sutskever]{radford2019language}
Radford, A., Wu, J., Child, R., Luan, D., Amodei, D., and Sutskever, I.
\newblock Language models are unsupervised multitask learners, 2019.

\bibitem[Rajbhandari et~al.(2019)Rajbhandari, Rasley, Ruwase, and
  He]{microsoftZero}
Rajbhandari, S., Rasley, J., Ruwase, O., and He, Y.
\newblock Zero: Memory optimizations toward training trillion parameter models,
  2019.
\newblock URL \url{https://arxiv.org/abs/1910.02054}.

\bibitem[Rajbhandari et~al.(2021)Rajbhandari, Ruwase, Rasley, Smith, and
  He]{microsoftZeroInfinity}
Rajbhandari, S., Ruwase, O., Rasley, J., Smith, S., and He, Y.
\newblock Zero-infinity: Breaking the gpu memory wall for extreme scale deep
  learning, 2021.
\newblock URL \url{https://arxiv.org/abs/2104.07857}.

\bibitem[Shallue et~al.(2018)Shallue, Lee, Antognini, Sohl-Dickstein, Frostig,
  and Dahl]{shallue2018}
Shallue, C.~J., Lee, J., Antognini, J., Sohl-Dickstein, J., Frostig, R., and
  Dahl, G.~E.
\newblock Measuring the effects of data parallelism on neural network training,
  2018.
\newblock URL \url{https://arxiv.org/abs/1811.03600}.

\bibitem[Shazeer et~al.(2018)Shazeer, Cheng, Parmar, Tran, Vaswani,
  Koanantakool, Hawkins, Lee, Hong, Young, et~al.]{googleMesh}
Shazeer, N., Cheng, Y., Parmar, N., Tran, D., Vaswani, A., Koanantakool, P.,
  Hawkins, P., Lee, H., Hong, M., Young, C., et~al.
\newblock Mesh-tensorflow: Deep learning for supercomputers.
\newblock \emph{Advances in neural information processing systems}, 31, 2018.

\bibitem[Shoeybi et~al.(2019)Shoeybi, Patwary, Puri, LeGresley, Casper, and
  Catanzaro]{nvidiaMegatron1}
Shoeybi, M., Patwary, M., Puri, R., LeGresley, P., Casper, J., and Catanzaro,
  B.
\newblock Megatron-lm: Training multi-billion parameter language models using
  model parallelism, 2019.
\newblock URL \url{https://arxiv.org/abs/1909.08053}.

\bibitem[Smith et~al.(2022)Smith, Patwary, Norick, LeGresley, Rajbhandari,
  Casper, Liu, Prabhumoye, Zerveas, Korthikanti, Zhang, Child, Aminabadi,
  Bernauer, Song, Shoeybi, He, Houston, Tiwary, and Catanzaro]{microsoftTuring}
Smith, S., Patwary, M., Norick, B., LeGresley, P., Rajbhandari, S., Casper, J.,
  Liu, Z., Prabhumoye, S., Zerveas, G., Korthikanti, V., Zhang, E., Child, R.,
  Aminabadi, R.~Y., Bernauer, J., Song, X., Shoeybi, M., He, Y., Houston, M.,
  Tiwary, S., and Catanzaro, B.
\newblock Using deepspeed and megatron to train megatron-turing nlg 530b, a
  large-scale generative language model, 2022.
\newblock URL \url{https://arxiv.org/abs/2201.11990}.

\bibitem[Smith et~al.(2018)Smith, Kindermans, and Le]{Smith2018DontDT}
Smith, S.~L., Kindermans, P.-J., and Le, Q.~V.
\newblock Don't decay the learning rate, increase the batch size.
\newblock \emph{ArXiv}, abs/1711.00489, 2018.

\bibitem[Vaswani et~al.(2017)Vaswani, Shazeer, Parmar, Uszkoreit, Jones, Gomez,
  Kaiser, and Polosukhin]{transformer}
Vaswani, A., Shazeer, N., Parmar, N., Uszkoreit, J., Jones, L., Gomez, A.~N.,
  Kaiser, L., and Polosukhin, I.
\newblock Attention is all you need, 2017.
\newblock URL \url{https://arxiv.org/abs/1706.03762}.

\bibitem[Zhang et~al.(2022)Zhang, Roller, Goyal, Artetxe, Chen, Chen, Dewan,
  Diab, Li, Lin, Mihaylov, Ott, Shleifer, Shuster, Simig, Koura, Sridhar, Wang,
  and Zettlemoyer]{facebookOpt}
Zhang, S., Roller, S., Goyal, N., Artetxe, M., Chen, M., Chen, S., Dewan, C.,
  Diab, M., Li, X., Lin, X.~V., Mihaylov, T., Ott, M., Shleifer, S., Shuster,
  K., Simig, D., Koura, P.~S., Sridhar, A., Wang, T., and Zettlemoyer, L.
\newblock Opt: Open pre-trained transformer language models, 2022.
\newblock URL \url{https://arxiv.org/abs/2205.01068}.

\end{thebibliography}

\appendix

\section{Analysis}
\label{sec:analysis}

\begin{table*}[tp]
{
\centering
\caption{Table of symbols.}
\label{tab:symbols}
\begin{tabular}{c|l|c}
\toprule
Symbol & Description & Formula \\
\midrule
\ngpu &  GPUs in the cluster & $\ndp\times\ntp\times\npp$\\ 
\ndp  & Data-parallel (\ddp) group size. Can be non-sharded (\nsdp), partially sharded (\psdp) \\ 
&  \quad or fully sharded (\fsdp)\\ 
\ntp & Tensor-parallel (\tp) group size & \\
\npp & Pipeline-parallel (\pp) group size. The schedule can be GPipe (\nlbf), 1F1B (\nldf), \\ 
& \quad breadth-first(\lbf), depth-first (\ldf) or Chimera\\ 
\bs & Batch size & $\ndp\times\mbs\times\mbc$ \\ 
\mbs  & Micro-batch size \\ 
\mbc  & Sequential micro-batches\\ 
\bcrit & Critical batch size \\ 
\bgpu & Batch size per GPU & $\bs/\ngpu$ \\ 
\bmin & Minimum batch size per GPU & $\ntp^{-1}$ \\ 
\btmin & Batch size per GPU for which $\tcomp=\tnet$ & $\approx\beta\tnet/\tcomp$ \\ 
\nl & Number of layers \\ 
\nstage & Number of pipeline stages. \\ 
\nloop & Number of pipeline loops & $\nstage/\npp$\\ 
\chimera & Number of pipelines in the Chimera schedule (even integer)\\ 
\tnet & Duration of a network operation (ex.: gradient reduction) \\ 
\tover & Duration of an overlapped computation (ex.: backward pass for the last micro-batch) \\ 
\tcomp & Total computation time (ex.: full backward pass)\\ 
\bottomrule
\end{tabular}
%
}\\
\end{table*}

We now turn to the theoretical analysis of large-scale distributed training, to provide a theoretical justification for the various assertions and claims made in earlier sections, and to demonstrate the benefits of Breadth-First Pipeline Parallelism. As already stated, distributed training is largely guided by memory, network, computational efficiency and batch size. We analyze each of these topics in isolation, then combine them to obtain general prescriptions.

\subsection{Setup}
\label{subsec:analysis_setup}

We begin by introducing the setup for our analysis. For clarity, we repeat some earlier definitions.

We consider a language model such as Bert \cite{bert} or GPT \cite{radford2019language}, with \nl identical transformer (encoder) layers \cite{transformer} with hidden size \hid. Such layers consist of a multi-head self-attention layer with \nhead heads of size \shead, followed by a two-layer MLP with hidden size \mlp. We assume the common choice $\nhead\times\shead=\hid$ and $\mlp=4\hid$. The model has a total of $\nparams\approx12\nl\hid^2$ parameters. We assume mixed precision training, the Adam optimizer  and activation checkpoints. 
Our analysis generalizes straightforwardly to other models and setups but may require extra considerations if the layers are not all identical.

The GPU \emph{cluster} consists of \nnode \emph{nodes} (servers) of \emph{size} \snode (typically 8), for a total of $\ngpu=\nnode\times\snode$ \emph{devices} (GPUs or similar machines such as TPUs). In modern Nvidia GPU clusters, the nodes are connected via an InfiniBand network, while the GPUs themselves are connected with a faster NVLink network.
When combining distributed methods, the cluster forms a (up to) three-dimensional grid $\ndp\times \ntp\times \npp$. The devices are parameterized by their \emph{ranks} (location on the grid), and the devices of constant pipeline and tensor rank form a self-contained data-parallel \emph{group} of \emph{size} \ndp, and so on. 
We designate the absence of a method with a group of size one.

With pipeline parallelism, the model is split into \nstage stages, 
looping $\nloop=\tfrac{\nstage}{\npp}$ times (one for non-looping pipelines, not necessarily an integer).
For a language model as considered here, each stage consists of a fixed number of transformer layers, while the input and output layers may form separate stages or be attached to others, whichever is preferable for a given scenario.

The input batch is split into \ndp parallel and \mbc sequential micro-batches of size \mbs, for a total batch size $\bs=\ndp\times \mbc\times \mbs$. The batch size per GPU \bgpu is minimized with $\mbc=\npp$ and $\mbs=1$, i.e., $\bgpu \geq \ntp^{-1}$. As $\ntp\leq\snode$, this implies $\bmin=\snode^{-1}$.
The bulk of the computation consists of matrix multiplications. These come from the linear layers, which require approximately 8 flop of computation per parameter and token, and from  self-attention, for a per-GPU total of
\begin{align}
    C_\text{GPU}&\approx\frac{96\mbc\mbs\nl\hid}{\npp\ntp}\nonumber\\
    &\qquad\times\left(\hid+\frac\seq6+\frac\voc{16\nl}\right).\label{eq:compute}\\
    &\approx\frac{96\mbc\mbs\nl\hid^2}{\npp\ntp}
    \label{eq:compute_approx}
\end{align}

To support our analysis, we consider two examples: GPT-3 ($\hid=12288$, $\nhead=\nl=96$) and a trillion-parameter model 1T ($\hid=12288$, $\nhead=160$, $\nl=128$), both trained with $\seq=2048$, $\ntp=8$. Unless otherwise specified, we also select a small pipeline with $\npp=4$

\subsection{Memory}
\label{subsec:memory}

The bulk of the memory usage falls in two main categories. First, the state memory consists of the training state and related variables such as half-precision buffers and parameter gradients, and scales proportionally with the model size. Second, the activation memory consists of the layer activations and their gradients, as well as the activation checkpoints. The activation memory scales principally with the input size, though it also scales with the model size.

\subsubsection{State memory}
\label{subsubsec:memory_state}

The state memory usage depends on the type of data parallelism used, and is approximately:
\begin{align}
    M_\nsdpi&= \frac{(12 \text{ to } 20)\nparams}{\npp\ntp},\\
    M_\psdpi&= \frac{(2 \text{ or } 4)\nparams}{\npp\ntp},\\
    M_\fsdpi&= \frac{8\nparams}{\nl\ntp}.
\end{align}
These formulas are justified as follows. For \nsdp, the bulk of the memory usage is from the training state itself, i.e., the full-precision weights and the optimizer momenta, which takes 12 bytes per parameter. The (full-precision) gradients and half-precision weight and gradient buffers may add up to 8 bytes per parameters, depending on the setup and implementation. With \psdp, the training state has a minimal memory usage given enough data parallelism, leaving the half-precision buffers as the main contributors to the state memory. With \lbf or $\mbc=1$, the gradients can be reduced immediately, dividing the memory usage by half. With \fsdp, the buffers are only required for the reconstructed layers. In general, two reconstructed layers are sufficient, which allows overlapping computation on a layer with reconstruction on another one.

For example, GPT-3 can be trained on 80 GB GPUs with $\ntp=8$ and $\npp=4$ using \psdp (10 or 20 GB), while 1T requires \fsdp (7 GB). Both models can also be trained with \nsdp but need larger pipelines with $\npp\geq8$ and $\npp\geq64$, respectively.


\subsubsection{Activation memory}
\label{subsubsec:memory_activation}

With activation checkpointing, the full activations and their gradients are only stored for one layer and micro-batch at the time. Their memory usage is approximated by \cite{nvidiaMegatron3}
\begin{equation}
    M_\text{act}=\seq\mbs\hid\left(10+\frac{24}{\ntp}+\frac{5\seq\nhead}{\hid\ntp}\right).
\end{equation}
This memory usage is minimal for a small micro-batch size and scales mildly with the model size. For example, GPT-3 uses 552 MB per sample, while 1T uses 1050 MB per sample. Note that due to memory fragmentation, the memory footprint may be significantly higher (see for example \cite{microsoftZero}).

For \nlbf or \lbf, the activation checkpoints have a memory usage of
\begin{equation}
    M_\text{ckpt}=\frac{\mbc\nl}{\npp}\times\frac{2\seq\mbs\hid}{\ntp}.
    \label{eq:mem_ckpt}
\end{equation}
For \nldf and \ldf, the number of checkpoints (first ratio) is capped to $(2\npp-1)\tfrac{\nl}{\npp}$ and $\nl+\npp-1$, respectively. When training at \bmin, the memory usage is 576 MB for GPT-3 and 1600 MB for 1T.

\subsection{Network}
\label{subsec:network}


As described in section \ref{sec:background}, the efficiency of network operations mainly depends on the ratio $\tcomp/\tnet$ of the compute and network times, and on the possibility of overlapping the two operations. This ratio can be estimated by comparing the \emph{arithmetic intensity} $I_\text{op}$ of the operations, defined as the ratio of computation performed with the amount of data transferred, with the ratio $I_\text{used}$ of compute and network that is actually performed per unit of time:
\begin{equation}
    \frac{\tcomp}{\tnet}=\frac{I_\text{op}}{I_\text{used}}
\end{equation}
Although $I_\text{used}$ is difficult to determine, it can be approximated by the known ratio $I_\text{hw}$ of available compute and network for the device, which we call the \emph{hardware intensity}:
\begin{equation}
    \frac{\tcomp}{\tnet}\approx\frac{I_\text{op}}{I_\text{hw}}
\end{equation}
For example, a NVidia A100 has 312 Tflop/s of available half-precision tensor core compute, and a network capacity of 46.6 GB/s with InfiniBand and 559 GB/s with NVLink,\footnote{These values differ slightly from the advertized values of 25 GB and 600 GB because of the conversion between base 10 and base 2, and because we consider the total bandwidth only (Input+Output).} resulting in intensities of $I_\text{NVLink}=520$ flop/byte and $I_\text{IB}=6240$ flop/byte.

\subsubsection{Data-parallel}
\label{subsubsec:network_data}

For \nsdp and \psdp, the network operations (reduction and reconstruction) transfer approximately 8 bytes per parameter per batch, which when compared to the computation (\ref{eq:compute_approx}), give an intensity of 
\begin{equation}
    I_\nsdpi\approx I_\psdpi=\mbc\mbs\seq.
\end{equation}
The intensity at \bmin is numerically equal to the sequence length. For example, when training on a A100 with $\seq=2048$, \btmin has the theoretical value $\lceil I_\text{op}/I_\text{IB}\rceil=4$. Note that the model size makes no difference.

With $\mbc>1$, only the breadth-first schedule allows overlapping with the whole batch.
The non-looped schedules can overlap with a single micro-batch, while depth-first is limited to \npp of them. This implies the following requirements for computational efficiency:
\begin{align}
    \text{Non-looped}&: \mbs\seq\geq I_\text{hw}\text{ or }\mbc\mbs\seq\gg I_\text{hw},\\
    \text{Depth-first}&: \npp\seq\geq I_\text{hw}\text{ or }\mbc\mbs\seq\gg I_\text{hw},\\
    \text{Breadth-first}&: \mbc\mbs\seq\geq I_\text{hw},
\end{align}
with the breadth-first case being far less constraining and potentially satisfied at \btmin.

With \fsdp, the repeated network operations reduce the intensity, depending on the schedule. With a non-looped pipeline or a non-pipelined schedule with standard gradient accumulation, the intensity becomes
\begin{equation}
    I_\fsdpi=\frac23\mbs\seq,
\end{equation}
in particular it is no longer affected by the micro-batch count.
Depth-first and breadth-first schedules improve this to 
\begin{align}
    I_{\fsdpi\text{-DF}}=\frac23\npp\mbs\seq,\\
    I_{\fsdpi\text{-BF}}=\frac23\mbc\mbs\seq.
\end{align}
The efficiency conditions become
\begin{align}
    \text{Non-looped}&: \frac23\mbs\seq\geq I_\text{hw},\\
    \text{Depth-first}&: \frac23\npp\seq\geq I_\text{hw},\\
    \text{Breadth-first}&: \frac23\mbc\mbs\seq\geq I_\text{hw},
\end{align}
i.e., a large micro-batch count no longer compensates for the poor overlap.

\subsubsection{Pipeline-parallel}
\label{subsubsec:network_pipe}

Pipeline parallelism requires about $4\tfrac{\hid}{\ntp\nl}$ bytes of network per token every $\tfrac\nl{\npp\nloop}$ layers, for an intensity
\begin{equation}
    I_\pp=24\hid\frac\nl{\npp\nloop}.
    \label{eq:intensity_pipeline}
\end{equation}
For $\npp=4$, this results in an intensity of 7.1 M for GPT-3 and 19.7 M for 1T when non-looped, or 294 K for GPT-3 and 614 K for 1T when maximally looped. All these values are far higher than the hardware intensities, but in practice the data transfers are much longer than predicted from Eq. (\ref{eq:intensity_pipeline}), and so is the overhead in the absence of overlap.

\subsubsection{Tensor-parallel}
\label{subsubsec:network_tensor}

In a transformer layer, tensor parallelism \cite{nvidiaMegatron1} requires approximately $96\frac{\hid^2}{\ntp}$ flop of computation and $48\hid$ bytes of network for each token, 2/3 of which cannot be overlapped,\footnote{The forward pass involves two non-overlapped all-reduce operations, each requiring 8 bytes of network for each hidden parameter and token. The backward pass adds an equivalent amount from the activation recomputation and two overlapped extra all-reduce from the gradient computation.
} for an arithmetic intensity
\begin{equation}
    I_\tp=2\frac{\hid}{\ntp}.
\end{equation}
As claimed, this  restricts \tp to the largest models and small-scale fast intra-node networks. For example, with $\ntp=8$, the intensity is 3072 for GPT-3 and 6400 for 1T, with expected overheads of about 11\% and 5\%, respectively.

\section{Critical batch size}
\label{sec:critical_batch_theory}

We provide a summary of the theoretical justification for Eq. (\ref{eq:critical_batch} )describing the overhead from the batch size. More details can be found in the original paper \cite{criticalBatch}.

In stochastic gradient descent, we attempt to minimize a loss function $L(\theta)$ having only access to a batch of noisy estimates $G_i$ of its gradient, $0\leq i<\bs$. By the central limit theorem, the average $G_\text{est}$ over the samples approximates to a multivariate normal distribution $\mathcal N(G(\theta),\Sigma(\theta))$, where $G(\theta)$ is the true gradient and the covariance matrix $\Sigma(\theta)$ is the noise. To the second order approximation, a step of size $-\epsilon G_\text{est}$ modifies the loss by
\begin{equation}
    \Delta L\approx -\epsilon G_\text{est}^T G+\frac12\epsilon^2G_\text{est}^T HG_\text{est},
\end{equation}
where $H$ is the Hessian, with expected value
\begin{equation}
    \mathbb E\left[\Delta L\right]\approx  -\epsilon |G|^2
    +\frac12\epsilon^2\left(G^THG+\text{tr}(H\Sigma)\right).
\end{equation}
This value is minimized with
\begin{equation}
    \epsilon=\frac{|G|^2}{G^THG+\text{tr}(H\Sigma)},\quad
    \mathbb E\left[\Delta L\right]\approx \frac{\tfrac12|G|^4}{G^THG+\text{tr}(H\Sigma)}.
    \label{eq:optimal_step}
\end{equation}
This result depends on the batch through the covariance matrix (from the central limit theorem). We extract that dependence by defining 
\begin{equation}
    \Sigma=\frac{\Sigma_0}B,\quad
    B_\text{noise}=\frac{\text{tr}(H\Sigma_0)}{G^THG}\approx\frac{\text{tr}(\Sigma_0)}{|G|^2}.
\end{equation}
The latter approximation assumes $H$ is close to the identity, which is not expected to hold in practice, but has been empirically shown to provide a fair estimate when the Hessian is unavailable.
Using these definitions, we rewrite Eq. (\ref{eq:optimal_step}) as
\begin{equation}
    \mathbb E\left[\Delta L\right]\approx \frac{\Delta L_0}{1+B_\text{noise}/\bs},
\end{equation}
where $\Delta L_0$ is an unimportant proportionality factor.
If neither of these quantities varies significantly, the same amount of progress is made each step, and a target loss is attained after a number of steps
\begin{equation}
    \text{Steps}\propto 1+\frac{B_\text{noise}}{\bs},
\end{equation}
In terms of samples seen, this rewrites as
\begin{equation}
    \text{Samples}\propto 1+\frac{\bs}{B_\text{noise}},
    \label{eq:critical_batch_noise}
\end{equation}
Despite the variety of assumptions and approximations used to obtain this results (second order, central limit theorem, optimal learning rate, consistent step, etc.), most of which are not expected to hold in practice, in \cite{criticalBatch} Eq. (\ref{eq:critical_batch_noise}) was shown to hold experimentally when replacing $B_\text{noise}$ by an empirical parameter \bcrit, generating Eq. (\ref{eq:critical_batch})
\begin{equation}
    \text{Samples}\propto 1+\frac{\bs}{\bcrit}.
\end{equation}
In most cases, $B_\text{noise}$ is a good approximation of the critical batch size, $\bcrit\approx B_\text{noise}$

\section{Breadth-first gradient accumulation}
\label{sec:breadth_first_2d}

\begin{figure*}[tp]
    \centering
    \subcaptionbox{
        Depth-first (\nsdp)
        \label{fig:schedule_df_2d}
    }{
        \hspace*{-8pt}\includegraphics[scale=1]{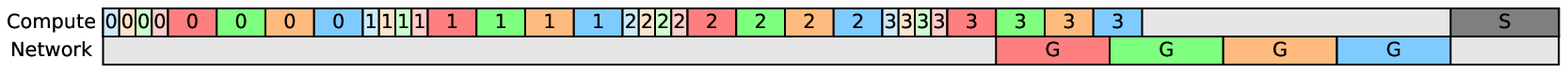}\vspace{-3pt}
    }\vspace{6pt}
    \subcaptionbox{
        Depth-first (\fsdp)
        \label{fig:schedule_df_sh_2d}
    }{
        \hspace*{-8pt}\includegraphics[scale=1]{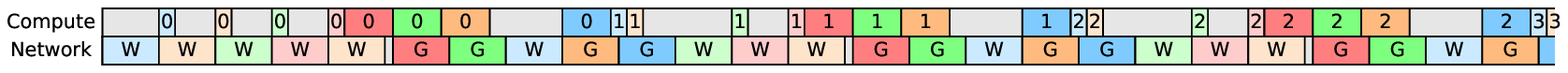}\vspace{-3pt}
    }\vspace{6pt}
    \subcaptionbox{
        Breadth-first (\nsdp)
        \label{fig:schedule_bf_2d}
    }{
        \hspace*{-8pt}\includegraphics[scale=1]{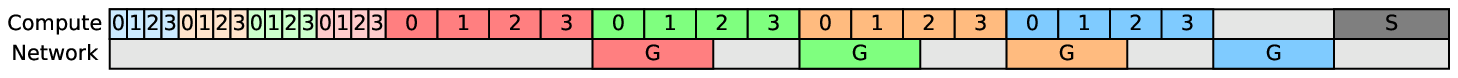}\vspace{-3pt}
    }\vspace{6pt}
    \subcaptionbox{
        Breadth-first (\fsdp)
        \label{fig:schedule_bf_sh_2d}
    }{
        \hspace*{-8pt}\includegraphics[scale=1]{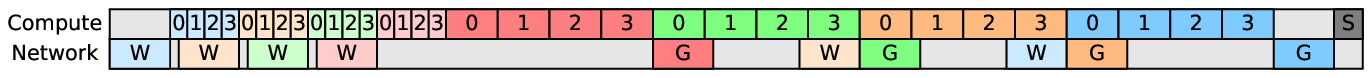}\vspace{-3pt}
    }\vspace{6pt}
    
    \includegraphics[trim={22pt 0 30pt 0}]{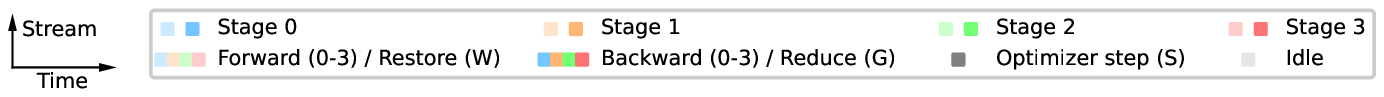}
    \caption{Example depth-first and breadth-first schedules for gradient accumulation with \nsdp and \fsdp. The depth-first approach achieves poor network overlap, and with \fsdp involves a costly repetition of the network operations. Both issues are solved with the breadth-first schedule, resulting in a faster training.}
    \label{fig:2d_acc_schedules}
\end{figure*}

We consider a data-parallel scenario with multiple sequential micro-batches. This may happen when a high batch size is needed to mitigate the gradient reduction network overhead, and the micro-batch size is limited by activation memory constraints. In that case, we typically use a \emph{depth-first} schedule, where a given micro-batch goes through the entire forward and backward passes before the next one begins. This schedule achieves the goal of limiting the memory usage, as all intermediate activations can be dropped between micro-batches. However, the gradient reduction cannot begin until the last micro-batch, leading to poor overlap with computation (Figure \ref{fig:schedule_bf_2d}). Therefore, the network overhead is mitigated, but not eliminated. With \fsdp there is no mitigation at all since the network operations (reconstruction and reduction) need to be repeated for each micro-batch (Figure \ref{fig:schedule_df_sh_2d}).

A breadth-first schedule solves these problem, allowing to overlap the gradient reduction with most of the backward pass, and with \fsdp avoiding duplicating the operations (Figures \ref{fig:schedule_bf_2d} and \ref{fig:schedule_bf_sh_2d}). The breadth-first schedule comes at the cost of memory, since more activations need to be stored at once. However, when using activation checkpoints, this memory increase remains small, and for larger models the memory savings from \fsdp are far more important.

However, when the stage outputs coincide with activation checkpoints, this only increases the checkpoint memory, which remains smaller than the layer activation unless there are lots of sequential micro-batches. Furthermore, for large models, the state memory is the bottleneck, so the memory usage may be \emph{lower} for the breadth-first schedule when combined with \fsdp.

\section{Implementation}
\label{sec:implementation}


We evaluated our method using a custom library, with a model and training scheme identical to the Megatron-LM implementation of Bert \cite{nvidiaMegatron1, nvidiaMegatron2}. As a reference, we used the source code for Megatron-LM as it was just before the publication of \cite{nvidiaMegatron3} (commit e156d2f). We verified that the model forward and backward passes are identical, generating the same kernels on the GPU, except for the fused GeLU (we used a slightly faster compiled implementation). However, our implementation differ for mixed precision (we overlap the casting), the optimization step (we removed some unnecessary synchronizations) and data and pipeline parallelism (as described below).

\subsection{Distributed training}
\label{subsec:distributed_implementation}

We used a custom implementation of data parallelism, with built-in support for mixed precision and sharded data parallelism. It relies on the breakdown into stages to optimize the data conversion and transfer, where the stage size may be set to any number of transformer layers. For that purpose, the embedding and final layers are either treated equivalently as transformer layers, or merged with the adjacent layer, depending on what is most efficient for a given configuration. We use a double-buffered approach to achieve network overlap at a minimal memory cost. For example, the computation for a given stage may be done in parallel with the weights for the next stage being restored on the other buffer. The network operations are performed in-place, avoiding the memory and kernel time overhead of network buffers. 

We implemented breadth-first pipeline parallelism, with support for network overlap as described in Section \ref{sec:breadth_first_pipeline}. It reduces to standard, non-looped pipeline parallelism when the stages are sufficiently large. 

\subsection{Memory efficiency}
\label{subsec:memory_efficiency}

Large models tend to suffer heavily from memory fragmentation, where the GPU has enough memory to allocate a given tensor but not in a contiguous chunk, which leads to unnecessary out-of-memory errors. To reduce the fragmentation, we pre-allocate tensor whenever possible, including for the training state (fp32 weights and gradients, optimizer momenta), fp16 weight and gradient buffers, the activation checkpoints, the pipeline receive buffers. Apart from a few small tensors, this leaves the intermediate layer activations and their gradients as the only actively allocated tensors; these still suffer heavily from memory fragmentation but are difficult to pre-allocate within Pytorch. 

We also observed a significant memory overhead and in some cases important slowdowns due to how the Pytorch caching allocator is implemented\footnote{This problem is not unique to our implementation. It was observed with \emph{Pytorch Fully Sharded Data Parallel}, 
and we were able to reproduce it on Megatron-LM without pipeline parallelism. (The Megatron-LM implementation of pipeline parallelism includes frequent CUDA synchronizations which, prevent the issue from happening, although inefficiently.)}. The tensor-parallel network operations are run in a separate CUDA stream set up by NCCL, and while that stream is immediately synchronized with the default stream, this limits the ability to free the tensors involved in the operation. As a result, the tensor memory is blocked from the CPU perspective until the operation is completed on GPU, which increases the memory usage when there are many queued kernels\footnote{In that scenario, some of the tensors involved may have been deleted on CPU, which means the underlying memory block will be available when the queued kernels complete. The CUDA caching allocator provides a way to reuse that memory before the kernels are run, by predicting the future memory usage. However, it can only do so efficiently in a single-stream setting.}. This may prevent the caching allocator from finding enough memory for future kernels, at which point it synchronizes the GPU, then flushes the cached allocations\footnote{The flushing is designed for a different scenario, where the memory is available but there is no cached block of the correct size. In the present case, the flush is generally unnecessary as the synchronization frees up many blocks but is performed either way.}. The flushing operation is relatively slow, causing some idle time on the GPU side. The overhead is multiplied when there are many parallel devices, as the slowdowns happen at different times, and each is enough to block the whole group. In some cases, we observed a combined overhead of more than 100\%. We fixed this by explicitly preventing the kernel queue from growing two big, by adding frequent CUDA synchronizations (on earlier events, so the synchronization is non-blocking.)

\section{Experimentation details and results}
\label{sec:detailed_results}

For each model, method and batch size, we ran a grid search over the following parameters, as applicable:
\begin{itemize}
    \item The implementation we used. This variable was fixed for all methods except non-looped, which was supported by both ours (GPipe) and Megatron-LM (1F1B).
    \item The pipeline-parallel group size \npp.
    \item The tensor-parallel group size \ntp.
    \item The micro-batch size \mbs.
    \item The number of sequential micro-batches \mbc.
    \item The number of transformer stages per device \nloop. This excludes the embedding and output layers which may add extra stages in our implementation.
    \item Whether we enabled \fsdp or \psdp. Note that we only tried \fsdp (not \psdp) for breadth-first and non-pipelined configurations and \psdp for non-looped configurations. We also did not try \psdp with Megatron-LM, which did not support it at the time. This may have led to an underestimation of the performance in some cases.
\end{itemize}
We excluded configurations that were obviously inferior, such as those with excessive model parallelism and those inefficiently combining \fsdp and gradient accumulation. We also excluded configurations that were certain or highly likely to run out of memory. 

We ran each configuration for 50 batches and measured the the average batch time, excluding the first 10 batches. We then calculated the throughput using Eq. (\ref{eq:compute}). We also measured the GPU memory usage, taken as the peak memory allocated on the GPU with global rank zero. This value is generally representative of the entire cluster, though the actual memory usage may vary slightly depending on the pipeline-parallel rank, and it does not take memory fragmentation into account.

Note that the measured memory usage does not reflect the true memory efficiency of each method because we optimized for throughput only, and because we used a relatively small cluster which limits the benefits of sharded data parallelism. To account for \fsdp and \psdp, we predicted the minimum memory usage achievable with with sharded data parallelism, i.e., on an arbitrarily large cluster. For our implementation, the difference is exactly 16 bits per parameter, while for Megatron-LM it is approximately 12 bits per parameter.\footnote{The difference is due to the full-precision gradients. They are pre-allocated in our implementation, which reduces memory fragmentation and mitigates the issues described in Appendix \ref{subsec:memory_efficiency}, but at the cost of extra memory. In Megatron-LM, they are instead allocated on the fly and do not typically contribute to the peak memory usage, which occurs around the beginning of the backward pass. Nevertheless, way be slightly overestimating the memory usage for smaller batch sizes where the activation memory is small, and the gradients potentially contribute to the peak memory.}


The most efficient configurations for each model, method and batch size are presented in tables \ref{tab:configs_52b}, \ref{tab:configs_6607m} and \ref{tab:configs_6607m_eth}, as well as the number of configurations tried in each case.\footnote{The number of configurations reflects the size of the configuration space and memory usage, and is not representative of the effort spent for each method. In particular, the larger number of configurations tried for the breadth-first schedule is due to its reduced memory usage (with \fsdp), which allows for many more distributed configurations.}

\begin{table*}[tp]
{
\centering
\caption{Selected optimal configurations for the 52 B model.}
\label{tab:configs_52b}
\scriptsize
\begin{tabular}{ccccccccccccc}
\toprule
 Method & Batch &  Implementation & \npp &  \ntp &  \mbs &  \mbc & \nloop &  Sharded & Throughput & Memory & Memory & Configs\\
 & size & &  &  &  &  &  & & (Tflop/s/GPU) & (GB) & min (GB) & \\
\midrule
\midrule
Breadth-first &           8 &               Ours &         8 &       8 &                 1 &                  8 &                              4 &                \xmark &              36.28 &                  15.96 &                               3.22 &     10 \\
Breadth-first &           9 &               Ours &         8 &       8 &                 1 &                  9 &                              8 &                \xmark &              42.33 &                  14.74 &                               2.25 &      3 \\
Breadth-first &          12 &               Ours &         8 &       8 &                 1 &                 12 &                              4 &                \xmark &              42.77 &                  16.66 &                               3.92 &     10 \\
Breadth-first &          16 &               Ours &         4 &       8 &                 1 &                  8 &                              8 &                \cmark &              44.49 &                  16.60 &                               3.60 &     30 \\
Breadth-first &          24 &               Ours &         4 &       8 &                 2 &                  6 &                              8 &                \cmark &              50.76 &                  17.96 &                               4.96 &     29 \\
Breadth-first &          32 &               Ours &         8 &       4 &                 1 &                 16 &                              4 &                \cmark &              51.17 &                  19.12 &                               5.13 &     59 \\
Breadth-first &          48 &               Ours &         8 &       2 &                 1 &                 12 &                              8 &                \cmark &              55.34 &                  19.73 &                               5.80 &     59 \\
Breadth-first &          64 &               Ours &         4 &       2 &                 1 &                  8 &                             16 &                \cmark &              54.01 &                  20.23 &                               6.30 &     89 \\
Breadth-first &         128 &               Ours &         4 &       2 &                 2 &                  8 &                              8 &                \cmark &              52.85 &                  24.65 &                              11.66 &     51 \\
Breadth-first &         256 &               Ours &         2 &       2 &                 1 &                 16 &                             32 &                \cmark &              48.97 &                  26.32 &                              12.38 &      5 \\
\midrule
  Depth-first &           8 &        Megatron-LM &         8 &       8 &                 1 &                  8 &                              2 &                \xmark &              29.53 &                  15.78 &                               6.42 &      3 \\
  Depth-first &          16 &        Megatron-LM &         8 &       8 &                 2 &                  8 &                              4 &                \xmark &              38.16 &                  15.94 &                               6.57 &      8 \\
  Depth-first &          24 &        Megatron-LM &         8 &       8 &                 1 &                 24 &                              2 &                \xmark &              38.37 &                  15.78 &                               6.42 &      3 \\
  Depth-first &          32 &        Megatron-LM &         8 &       8 &                 4 &                  8 &                              4 &                \xmark &              43.50 &                  17.77 &                               8.41 &     13 \\
  Depth-first &          48 &        Megatron-LM &         8 &       8 &                 2 &                 24 &                              2 &                \xmark &              45.52 &                  16.27 &                               6.91 &      8 \\
  Depth-first &          64 &        Megatron-LM &         8 &       8 &                 4 &                 16 &                              4 &                \xmark &              48.52 &                  19.18 &                               9.81 &     15 \\
  Depth-first &         128 &        Megatron-LM &         8 &       8 &                 4 &                 32 &                              4 &                \xmark &              51.46 &                  19.18 &                               9.81 &     18 \\
  Depth-first &         256 &        Megatron-LM &        16 &       4 &                 4 &                 64 &                              2 &                \xmark &              54.91 &                  21.35 &                              11.62 &     19 \\
  Depth-first &         512 &        Megatron-LM &         8 &       8 &                 4 &                128 &                              2 &                \xmark &              55.41 &                  19.87 &                              10.50 &      8 \\
\midrule
   Non-looped &           8 &               Ours &         8 &       8 &                 1 &                  8 &                              1 &                \xmark &              26.04 &                  16.87 &                               4.38 &      3 \\
   Non-looped &           9 &               Ours &         8 &       8 &                 1 &                  9 &                              1 &                \xmark &              27.59 &                  16.99 &                               4.50 &      1 \\
   Non-looped &          12 &               Ours &         8 &       8 &                 1 &                 12 &                              1 &                \xmark &              30.74 &                  17.38 &                               4.89 &      2 \\
   Non-looped &          16 &               Ours &         8 &       8 &                 1 &                 16 &                              1 &                \xmark &              33.54 &                  17.89 &                               5.40 &      9 \\
   Non-looped &          24 &               Ours &         8 &       8 &                 1 &                 24 &                              1 &                \xmark &              37.46 &                  18.91 &                               6.42 &      7 \\
   Non-looped &          32 &               Ours &         8 &       8 &                 2 &                 16 &                              1 &                \xmark &              39.62 &                  20.12 &                               7.63 &     16 \\
   Non-looped &          48 &               Ours &         8 &       4 &                 1 &                 24 &                              1 &                \cmark &              44.30 &                  22.71 &                               9.74 &     14 \\
   Non-looped &          64 &               Ours &         8 &       4 &                 1 &                 32 &                              1 &                \cmark &              46.74 &                  23.75 &                              10.78 &     19 \\
   Non-looped &         128 &        Megatron-LM &         8 &       8 &                 2 &                 64 &                              1 &                \xmark &              49.35 &                  15.75 &                               6.38 &     10 \\
   Non-looped &         256 &        Megatron-LM &        16 &       4 &                 2 &                128 &                              1 &                \xmark &              53.72 &                  16.33 &                               6.61 &      8 \\
   Non-looped &         512 &        Megatron-LM &         8 &       8 &                 4 &                128 &                              1 &                \xmark &              55.52 &                  17.68 &                               8.31 &      4 \\
\midrule
  No pipeline &           8 &               Ours &         1 &       8 &                 1 &                  1 &                              1 &                \cmark &               4.73 &                  14.23 &                               1.98 &      1 \\
  No pipeline &          16 &               Ours &         1 &       8 &                 2 &                  1 &                              1 &                \cmark &               9.43 &                  15.44 &                               3.19 &      3 \\
  No pipeline &          24 &               Ours &         1 &       8 &                 3 &                  1 &                              1 &                \cmark &              14.07 &                  16.64 &                               4.39 &      1 \\
  No pipeline &          32 &               Ours &         1 &       8 &                 4 &                  1 &                              1 &                \cmark &              18.79 &                  17.85 &                               5.60 &      6 \\
  No pipeline &          48 &               Ours &         1 &       8 &                 6 &                  1 &                              1 &                \cmark &              27.66 &                  20.29 &                               8.04 &      3 \\
  No pipeline &          64 &               Ours &         1 &       8 &                 8 &                  1 &                              1 &                \cmark &              35.97 &                  22.73 &                              10.48 &     10 \\
  No pipeline &         128 &               Ours &         1 &       2 &                 4 &                  1 &                              1 &                \cmark &              53.01 &                  21.43 &                               9.18 &     12 \\
  No pipeline &         256 &               Ours &         1 &       2 &                 4 &                  2 &                              1 &                \cmark &              51.57 &                  21.43 &                               9.18 &     12 \\
  No pipeline &         512 &               Ours &         1 &       2 &                 4 &                  4 &                              1 &                \cmark &              62.40 &                  21.44 &                               9.19 &      7 \\
\bottomrule
\end{tabular}

}

\end{table*}

\begin{table*}[tp]
{
\centering
\caption{Selected optimal configurations for the 6.6 B model.}
\label{tab:configs_6607m}
\scriptsize
\begin{tabular}{ccccccccccccc}
\toprule
 Method & Batch &  Implementation & \npp &  \ntp &  \mbs &  \mbc & \nloop &  Sharded & Throughput & Memory & Memory & Configs\\
 & size & &  &  &  &  &  & & (Tflop/s/GPU) & (GB) & min (GB) & \\
\midrule
Breadth-first &          32 &               Ours &         4 &       2 &                 1 &                  4 &                              4 &                \xmark &              35.72 &                  15.56 &                               2.34 &     15 \\
Breadth-first &          48 &               Ours &         4 &       2 &                 1 &                  6 &                              8 &                \xmark &              43.66 &                  14.61 &                               1.64 &     15 \\
Breadth-first &          64 &               Ours &         2 &       2 &                 1 &                  4 &                              4 &                \cmark &              46.60 &                   5.67 &                               4.02 &     35 \\
Breadth-first &          96 &               Ours &         2 &       1 &                 1 &                  3 &                              8 &                \cmark &              54.07 &                   5.95 &                               4.30 &     35 \\
Breadth-first &         128 &               Ours &         2 &       1 &                 1 &                  4 &                              8 &                \cmark &              57.03 &                   6.10 &                               4.45 &     55 \\
Breadth-first &         192 &               Ours &         2 &       1 &                 2 &                  3 &                              8 &                \cmark &              59.55 &                   6.72 &                               5.06 &     55 \\
Breadth-first &         256 &               Ours &         2 &       1 &                 2 &                  4 &                              8 &                \cmark &              60.45 &                   7.02 &                               5.36 &     71 \\
Breadth-first &         384 &               Ours &         2 &       1 &                 4 &                  3 &                              8 &                \cmark &              59.02 &                   8.25 &                               6.59 &     71 \\
Breadth-first &         512 &               Ours &         2 &       1 &                 4 &                  4 &                             16 &                \cmark &              57.44 &                   8.95 &                               5.52 &     80 \\
 \midrule
 Depth-first &          32 &        Megatron-LM &         4 &       2 &                 1 &                  4 &                              2 &                \xmark &              27.27 &                  16.27 &                               6.54 &      3 \\
  Depth-first &          64 &        Megatron-LM &         4 &       2 &                 2 &                  4 &                              4 &                \xmark &              35.24 &                  16.35 &                               6.62 &      8 \\
  Depth-first &          96 &        Megatron-LM &         4 &       2 &                 1 &                 12 &                              2 &                \xmark &              38.00 &                  16.27 &                               6.54 &      3 \\
  Depth-first &         128 &        Megatron-LM &         4 &       2 &                 4 &                  4 &                              4 &                \xmark &              42.33 &                  16.44 &                               6.72 &     13 \\
  Depth-first &         192 &        Megatron-LM &         4 &       2 &                 2 &                 12 &                              2 &                \xmark &              45.55 &                  16.29 &                               6.56 &      8 \\
  Depth-first &         256 &        Megatron-LM &         4 &       2 &                 4 &                  8 &                              4 &                \xmark &              47.89 &                  16.44 &                               6.72 &     18 \\
  Depth-first &         384 &        Megatron-LM &         4 &       2 &                 4 &                 12 &                              2 &                \xmark &              50.14 &                  16.32 &                               6.59 &     13 \\
  Depth-first &         512 &        Megatron-LM &         4 &       2 &                 4 &                 16 &                              2 &                \xmark &              51.92 &                  16.32 &                               6.59 &     20 \\
 \midrule
  Non-looped &          32 &               Ours &         4 &       2 &                 1 &                  4 &                              1 &                \xmark &              25.42 &                  16.73 &                               3.76 &      4 \\
   Non-looped &          48 &               Ours &         4 &       2 &                 1 &                  6 &                              1 &                \xmark &              30.88 &                  16.86 &                               3.89 &      2 \\
   Non-looped &          64 &               Ours &         4 &       2 &                 1 &                  8 &                              1 &                \xmark &              34.63 &                  17.00 &                               4.03 &     10 \\
   Non-looped &          96 &               Ours &         4 &       2 &                 1 &                 12 &                              1 &                \xmark &              39.13 &                  17.27 &                               4.30 &      7 \\
   Non-looped &         128 &               Ours &         4 &       2 &                 1 &                 16 &                              1 &                \xmark &              41.72 &                  17.54 &                               4.57 &     16 \\
   Non-looped &         192 &               Ours &         4 &       1 &                 1 &                 12 &                              1 &                \cmark &              47.09 &                  11.21 &                               7.78 &     12 \\
   Non-looped &         256 &               Ours &         4 &       1 &                 1 &                 16 &                              1 &                \cmark &              50.25 &                  11.49 &                               8.06 &     21 \\
   Non-looped &         384 &               Ours &         4 &       1 &                 1 &                 24 &                              1 &                \cmark &              52.90 &                  12.06 &                               8.63 &     19 \\
   Non-looped &         512 &               Ours &         4 &       1 &                 2 &                 16 &                              1 &                \cmark &              52.78 &                  12.94 &                               9.51 &     24 \\
 \midrule
 No pipeline &          32 &               Ours &         1 &       8 &                 4 &                  1 &                              1 &                \xmark &              23.19 &                  14.04 &                               1.72 &      6 \\
  No pipeline &          48 &               Ours &         1 &       8 &                 6 &                  1 &                              1 &                \xmark &              32.64 &                  14.74 &                               2.42 &      3 \\
  No pipeline &          64 &               Ours &         1 &       8 &                 8 &                  1 &                              1 &                \xmark &              40.73 &                  15.45 &                               3.13 &     10 \\
  No pipeline &          96 &               Ours &         1 &       8 &                12 &                  1 &                              1 &                \xmark &              47.44 &                  16.89 &                               4.57 &      6 \\
  No pipeline &         128 &               Ours &         1 &       2 &                 4 &                  1 &                              1 &                \cmark &              55.73 &                   4.40 &                               2.82 &     13 \\
  No pipeline &         192 &               Ours &         1 &       2 &                 6 &                  1 &                              1 &                \cmark &              57.74 &                   5.30 &                               3.72 &     10 \\
  No pipeline &         256 &               Ours &         1 &       1 &                 4 &                  1 &                              1 &                \cmark &              60.02 &                   6.01 &                               4.43 &     15 \\
  No pipeline &         384 &               Ours &         1 &       1 &                 6 &                  1 &                              1 &                \cmark &              59.45 &                   7.19 &                               5.62 &     13 \\
  No pipeline &         512 &               Ours &         1 &       1 &                 8 &                  1 &                              1 &                \cmark &              59.01 &                   8.38 &                               6.80 &     16 \\
\bottomrule
\end{tabular}

}
\end{table*}

\begin{table*}[tp]
{
\centering
\caption{Selected optimal configurations for the 6.6 B model (Ethernet).}
\label{tab:configs_6607m_eth}
\scriptsize
\begin{tabular}{ccccccccccccc}
\toprule
 Method & Batch &  Implementation & \npp &  \ntp &  \mbs &  \mbc & \nloop &  Sharded & Throughput & Memory & Memory & Configs\\
 & size & &  &  &  &  &  & & (Tflop/s/GPU) & (GB) & min (GB) & \\
\midrule
Breadth-first &          64 &               Ours &         4 &       4 &                 2 &                  8 &                              4 &                \xmark &              31.31 &                   8.70 &                               2.21 &     88 \\
Breadth-first &          96 &               Ours &         4 &       4 &                 4 &                  6 &                              4 &                \xmark &              36.31 &                   9.47 &                               2.98 &     88 \\
Breadth-first &         128 &               Ours &         2 &       4 &                 4 &                  4 &                              8 &                \xmark &              40.00 &                  16.40 &                               3.79 &    113 \\
Breadth-first &         192 &               Ours &         2 &       4 &                 8 &                  3 &                              8 &                \xmark &              47.44 &                  18.04 &                               5.43 &    113 \\
Breadth-first &         256 &               Ours &         2 &       4 &                 4 &                  8 &                              8 &                \xmark &              46.85 &                  18.83 &                               6.21 &    121 \\
Breadth-first &         384 &               Ours &         2 &       4 &                16 &                  3 &                              4 &                \xmark &              50.86 &                  23.35 &                              10.73 &    130 \\
Breadth-first &         512 &               Ours &         2 &       4 &                16 &                  4 &                              8 &                \xmark &              50.80 &                  25.02 &                              12.41 &    106 \\
\midrule
  Depth-first &          64 &        Megatron-LM &         8 &       2 &                 2 &                  8 &                              2 &                \xmark &              25.40 &                   8.78 &                               3.56 &     25 \\
  Depth-first &          96 &        Megatron-LM &         8 &       2 &                 1 &                 24 &                              2 &                \xmark &              26.94 &                   8.77 &                               3.54 &     16 \\
  Depth-first &         128 &        Megatron-LM &         8 &       1 &                 1 &                 16 &                              2 &                \xmark &              31.28 &                  17.43 &                               6.98 &     28 \\
  Depth-first &         192 &        Megatron-LM &         8 &       1 &                 1 &                 24 &                              2 &                \xmark &              33.70 &                  17.43 &                               6.98 &     25 \\
  Depth-first &         256 &        Megatron-LM &         8 &       1 &                 2 &                 16 &                              2 &                \xmark &              36.37 &                  17.45 &                               7.00 &     28 \\
  Depth-first &         384 &        Megatron-LM &         8 &       1 &                 2 &                 24 &                              2 &                \xmark &              39.24 &                  17.45 &                               7.00 &     28 \\
  Depth-first &         512 &        Megatron-LM &         8 &       1 &                 2 &                 32 &                              2 &                \xmark &              40.75 &                  17.45 &                               7.00 &     28 \\
 \midrule
  Non-looped &          64 &               Ours &         8 &       2 &                 1 &                 16 &                              1 &                \xmark &              29.70 &                   9.52 &                               2.55 &     40 \\
   Non-looped &          96 &               Ours &         8 &       2 &                 1 &                 24 &                              1 &                \xmark &              33.91 &                   9.81 &                               2.84 &     35 \\
   Non-looped &         128 &               Ours &         8 &       2 &                 1 &                 32 &                              1 &                \xmark &              36.05 &                  10.10 &                               3.13 &     52 \\
   Non-looped &         192 &               Ours &         8 &       1 &                 1 &                 24 &                              1 &                \xmark &              40.42 &                  18.78 &                               4.85 &     50 \\
   Non-looped &         256 &               Ours &         8 &       1 &                 1 &                 32 &                              1 &                \xmark &              43.66 &                  19.10 &                               5.17 &     56 \\
   Non-looped &         384 &               Ours &         8 &       1 &                 1 &                 48 &                              1 &                \xmark &              47.58 &                  19.74 &                               5.81 &     60 \\
   Non-looped &         512 &               Ours &         8 &       1 &                 1 &                 64 &                              1 &                \xmark &              48.48 &                  20.38 &                               6.45 &     49 \\
\midrule
  No pipeline &          64 &               Ours &         1 &       8 &                 8 &                  1 &                              1 &                \xmark &              15.37 &                  15.45 &                               3.13 &      4 \\
  No pipeline &          96 &               Ours &         1 &       8 &                12 &                  1 &                              1 &                \xmark &              21.43 &                  16.89 &                               4.57 &      3 \\
  No pipeline &         128 &               Ours &         1 &       8 &                16 &                  1 &                              1 &                \xmark &              27.65 &                  18.33 &                               6.02 &      5 \\
  No pipeline &         192 &               Ours &         1 &       8 &                24 &                  1 &                              1 &                \xmark &              37.35 &                  21.22 &                               8.90 &      4 \\
  No pipeline &         256 &               Ours &         1 &       8 &                32 &                  1 &                              1 &                \xmark &              45.99 &                  24.10 &                              11.78 &      5 \\
  No pipeline &         384 &               Ours &         1 &       8 &                48 &                  1 &                              1 &                \cmark &              46.81 &                  19.09 &                              17.51 &      5 \\
  No pipeline &         512 &               Ours &         1 &       8 &                32 &                  2 &                              1 &                \xmark &              46.40 &                  24.13 &                              11.81 &      5 \\
\bottomrule
\end{tabular}

}
\end{table*}

\end{document}